\documentclass[twocolumn,twoside,10pt,journal]{IEEEtran}

\usepackage{epsfig,amsfonts,color,amsbsy}
\usepackage[nolist]{acronym}
\usepackage{graphicx,cite,amssymb,amsthm,bm}
\usepackage{amsmath}
\usepackage{mathrsfs}
\usepackage{bm,bbm}
\usepackage{multirow}
\usepackage{bigstrut}
\usepackage{psfrag}
\usepackage{stackengine}
\usepackage{xcolor}
\usepackage{mathtools}
\usepackage{tabularx,array}
\usepackage{mathbbol}
\usepackage{lipsum}
\usepackage[sans]{dsfont}
\mathtoolsset{showonlyrefs}
\usepackage{array, makecell}
\usepackage{float}
\usepackage{balance}
\usepackage{verbatim}
\usepackage{mathrsfs}
\usepackage{tikz}
\usepackage{pgfplots}
\usepackage{scalerel}
\usepackage[caption=false]{subfig}

\pgfplotsset{compat=1.17}

\usepackage{tikz}
\usetikzlibrary{shapes,arrows}
\usetikzlibrary{calc}
\newtheorem{example}{Example}

\definecolor{TUMBeamerYellow}    {rgb}{1.00,0.71,0.00}  % RGB 255,180,000
\definecolor{TUMBeamerOrange}    {rgb}{1.00,0.50,0.00}  % RGB 255,128,000
\definecolor{TUMBeamerRed}       {rgb}{0.90,0.20,0.09}  % RGB 229,052,024
\definecolor{TUMBeamerBlue}      {rgb}{0.00,0.60,1.00}  % RGB 000,153,255
\definecolor{TUMBeamerLightBlue} {rgb}{0.25,0.75,1.00}  % RGB 065,190,255
\definecolor{TUMBeamerGreen}     {rgb}{0.57,0.67,0.42}  % RGB 145,172,107

\DeclareMathOperator*{\argmax}{arg\,max}

\newtheorem{lemma}{Lemma}
\newtheorem{definition}{Definition}
\newtheorem{theorem}{Theorem}

\newtheorem{remark}{Remark}

\newcommand{\bmatxcap}[2]{\bm{B}_{\scaleto{#2}{4pt}}^{\scaleto{(#1)}{4pt}}}
\newcommand{\bmatx}[2]{\bm{B}_{\scaleto{#2}{6pt}}^{\scaleto{(#1)}{5pt}}}
\newcommand{\bmatonehalfx}{\bm{B}_{\scaleto{1/2}{6pt}}}

\newcommand{\sscr}{s}
\newcommand{\Sscr}{S}
\newcommand{\vecscr}{\bm{\sscr}}
\newcommand{\vecScr}{\bm{\Sscr}}

\newcommand{\WCL}{\mathsf{WCL}}

\newcommand{\rmin}{\underline{R}}
\newcommand{\rmax}{\overline{R}}
\newcommand{\omin}{\underline{\omega}}
\newcommand{\omax}{\overline{\omega}}
\newcommand{\rset}{\mathcal{R}}

\newcommand{\thr}{\gamma^\star}

\newcommand{\inR}{R_{\scaleto{\mathsf{I}}{3.5pt}}}
\newcommand{\inmR}{R_{\scaleto{\mathsf{IM}}{3.5pt}}}
\newcommand{\dmR}{R_{\scaleto{\mathsf{O}}{3.5pt}}}
\newcommand{\Hb}{\mathsf{H}_{\scaleto{\mathsf{b}}{4pt}}}
\newcommand{\incode}{\mathcal{C}_{\scaleto{\mathsf{I}}{3.5pt}}}
\newcommand{\inmcode}{\mathcal{C}_{\scaleto{\mathsf{IM}}{3.5pt}}}
\newcommand{\outcode}{\mathcal{C}_{\scaleto{\mathsf{O}}{3.5pt}}}
\newcommand{\code}{\mathcal{C}}
\newcommand{\PEP}{\mathsf{PEP}}
\newcommand{\prob}[1]{\mathsf{P}\!\left[#1\right]}
\newcommand{\expect}[1]{\mathsf{E}\!\left[#1\right]}
\newcommand{\iowe}{A^{\scaleto{\mathsf{IO}}{3.5pt}}}
\newcommand{\aiowe}{\bar{A}^{\scaleto{\mathsf{IO}}{3.5pt}}}
\newcommand{\fieldtwo}{\mathbbmss{F}_{\!\scaleto{\mathsf{2}}{3.5pt}}}
\newcommand{\liftfac}{\ell}
\newcommand{\hw}{w_{\scaleto{\mathsf{H}}{3.5pt}}}
\newcommand{\hd}{d_{\scaleto{\mathsf{H}}{3.5pt}}}

\newcommand{\cbsc}{C_{\scaleto{\mathsf{BSC}}{3.5pt}}}
\newcommand{\cawgn}{C_{\scaleto{\mathsf{AWGN}}{3.5pt}}}

\newcommand{\vecv}{\bm{v}}
\newcommand{\vecx}{\bm{x}}
\newcommand{\vecy}{\bm{y}}
\newcommand{\vecw}{\bm{w}}
\newcommand{\vecc}{\bm{c}}

\newcommand{\vecb}{\bm{b}}

\newcommand{\vecY}{\bm{Y}}

\newcommand{\ensP}[2]{\mathscr{C}_{#2}\!\left(#1\right)}
\newcommand{\ensfP}[1]{\mathscr{C}\!\left(#1\right)}
\newcommand{\pgraph}{\mathcal{P}}
\newcommand{\bgraph}{\mathcal{G}}

\newcommand{\vn}{\mathsf{v}}
\newcommand{\cn}{\mathsf{c}}
\newcommand{\transposed}{{\scaleto{\mathsf{T}}{3.5pt}}}

\newcommand{\vB}{\bm{B}}
\newcommand{\vH}{\bm{H}}
\newcommand{\vX}{\bm{X}}

\newcommand{\xML}{\hat{\vecx}_{\scaleto{\mathsf{ML}}{3.5pt}}}
\newcommand{\xMM}{\hat{\vecx}_{\scaleto{\mathsf{MM}}{3.5pt}}}

\newcommand{\dci}{\mathsf{d}_{\cn_{i}}}

\newcommand{\dvj}{\mathsf{d}_{\vn_{j}}}

\newcommand{\edges}{\mathscr{E}}
\newcommand{\edgesci}{\edges_{\cn_i}}
\newcommand{\edgesvj}{\edges_{\vn_j}}
\newcommand{\weightv}{\bm{\epsilon}}
\newcommand{\weightvj}{\epsilon}
\newcommand{\ttheta}{\tilde{\weightvj}}
\newcommand{\setVN}{\mathcal{I}} % set of VNs
\newcommand{\kp}{\kappa}
\newcommand{\tk}{\tilde{\kp}}
\newcommand{\cV}{\mathcal{V}}
\newcommand{\cC}{\mathcal{C}}
\DeclareMathOperator{\coeff}{coeff}
\newcommand{\wH}{\mathsf{w}_{\text{H}}}

\IEEEoverridecommandlockouts
\begin{document}

\title{Rate-Adaptive Protograph-Based\\ MacKay-Neal Codes}

\author{
	Ayman Zahr, \IEEEmembership{Graduate Student Member, IEEE}, Emna Ben Yacoub, \IEEEmembership{Graduate Student Member, IEEE}, Bal\'azs Matuz, \IEEEmembership{Senior Member, IEEE}, Gianluigi Liva \IEEEmembership{Senior Member, IEEE}
	\thanks{This work was presented in part at the IEEE Information Theory Workshop, Saint Malo, France, April 2023 \cite{Zahr2023}.
	}
	\thanks{
		A. Zahr is with the Institute for Communications Engineering, School of Computation, Information, and Technology, Technical University of Munich, Arcisstrasse 21, Munich, Germany, and with the Institute of Communications and Navigation, German Aerospace Center (DLR), Wessling, Germany, (email:  ayman.zahr@dlr.de).
	}
	\thanks{
	G. Liva is  with the Institute of Communications and Navigation, German Aerospace Center (DLR), Wessling, Germany (email:  gianluigi.liva@dlr.de).
	}
	\thanks{
		E. Ben Yacoub and B. Matuz are with the Huawei Munich Research Center, 80992 Munich, Germany (email: \{emna.ben.yacoub,balazs.matuz\}@huawei.com).
	}
	\thanks{G. Liva  acknowledges the financial support by the Federal Ministry of Education and Research of Germany in the programme of "Souver\"an. Digital. Vernetzt." Joint project 6G-RIC, project identification number: 16KISK022.}
	}

\maketitle
\thispagestyle{empty}
\IEEEoverridecommandlockouts

%\markboth
%{IEEE Transactions on Information Theory}
%{A. Zahr et al.: Rate-Adaptive Protograph MN Codes}

\begin{abstract}
	Rate-adaptive MacKay-Neal (MN) codes based on protographs are analyzed.  The code construction employs an outer distribution matcher (DM) to adapt the rate of the scheme. The DM is coupled with an inner protograph-based low-density parity-check (LDPC) code. The performance achievable by the resulting code structure, that is nonlinear, is studied by means of an equivalent communication model that reduces the problem to the analysis of the inner (linear) LDPC code with transmission that takes place in parallel over the communication channel, and over a suitably defined binary symmetric channel. A density evolution analysis of protograph MN code ensembles is outlined, and it is complemented by an error floor analysis that relies on the derivation of the average input-output weight distribution of the inner LDPC code ensemble. Conditions on the shape of the normalized logarithmic asymptotic input-output weight distribution are defined, which allow discarding code ensembles with bad error floor properties during the code design phase. Examples of code designs are provided, showing how the use of a single LDPC code ensemble allows operating within $1$ dB from the Shannon limit over a wide range of code rates, where the code rate is selected by tuning the DM parameters. By enabling rate flexibility with a constant blocklength, and with a fixed LDPC code as inner code, the construction provides an appealing solution for very high-throughput wireless (optical) links that employ binary-input modulations. 
\end{abstract}

\pagenumbering{arabic}

%%%%%%%%%%%%%%%%%%%%%%%%%%%%%%%%%%%%%%%%%%%%%%%%%%%%%%%%%%%%%%%%%%%%%%%%%
%%%%%%%%%%%%%%%%%%%%%%%%%%%%%%%%%%%%%%%%%%%%%%%%%%%%%%%%%%%%%%%%%%%%%%%%%
%%%%%%%%%%%%%%%%%%%%%%%%%%%%%%%%%%%%%%%%%%%%%%%%%%%%%%%%%%%%%%%%%%%%%%%%%

\begin{acronym}
	\acro{BEC}{binary erasure channel}
	\acro{BP}{belief propagation}
	\acro{DE}{density evolution}
	\acro{LDPC}{low-density parity-check}
	\acro{ML}{maximum likelihood}
	\acro{r.v.}{random variable}
	\acro{PEP}{pairwise error probability}
	\acro{BP}{belief propagation}
	\acro{BPSK}{binary phase shift keying}
	\acro{BSC}{binary symmetric channel}
	\acro{AWGN}{additive white Gaussian noise}
	\acro{OOK}{on-off keying}
	\acro{DM}{distribution matcher}
	\acro{p.m.f.}{probability mass function}
	\acro{p.d.f.}{probability density function}
	\acro{i.i.d.}{independent and identically-distributed}
	\acro{CC}{constant composition}
	\acro{LEO}{low earth orbit}
	\acro{biAWGN}{binary-input additive white Gaussian noise}
	\acro{PAM}{pulse amplitude modulation}
	\acro{SNR}{signal-to-noise ratio}
	\acro{EXIT}{extrinsic information transfer}
	\acro{PEXIT}{protograph extrinsic information transfer}
	\acro{VN}{variable node}
	\acro{CN}{check node}
	\acro{FER}{frame error rate}
	\acro{MN}{MacKay-Neal}
	\acro{RA}{repeat-accumulate}
	\acro{NS}{non-systematic}
	\acro{LLR}{log-likelihood ratio}
	\acro{MM}{mismatched}
	\acro{MBIOS}{memoryless binary-input output-symmetric} 
	\acro{EPC}{\emph{equivalent parallel channel}}
	\acro{QPSK}{quadrature phase shift keying}
	\acro{WCL}{worst-case loss}
	\acro{SoF}{start-of-frame}
	\acro{SC}{spatially coupled}
	\acro{UB}{union bound}
	\acro{TUB}{truncated union bound}
	\acro{PEG}{progressive edge growth}
	\acro{PAS}{probabilistic amplitude shaping}
	\acro{CCDM}{constant composition distribution matcher}
	\acro{MI}{mutual information}
\end{acronym}

%%%%%%%%%%%%%%%%%%%%%%%%%%%%%%%%%%%%%%%%%%%%%%%%%%%%%%%%%%%%%%%%%%%%%%%%%
%%%%%%%%%%%%%%%%%%%%%%%%%%%%%%%%%%%%%%%%%%%%%%%%%%%%%%%%%%%%%%%%%%%%%%%%%
%%%%%%%%%%%%%%%%%%%%%%%%%%%%%%%%%%%%%%%%%%%%%%%%%%%%%%%%%%%%%%%%%%%%%%%%%

\section{Introduction}\label{sec:intro}

\IEEEPARstart{T}{he} development of high-speed communication links, either in the fiber optics or in the wireless (radio/optical) domains, calls for the development of channel codes that support fast decoding algorithms \cite{Smith2012a,Pfi17}. For data rates in the order of several tens of Gbps, some key techniques are currently considered to enable high-speed decoding. On the algorithmic side, the use of (generalized) \ac{LDPC} \cite{Gal63,Tan81} codes with hard-decision message passing decoders \cite{Smith2012a,Pfi17,Lechner2012,Zha18} has been recently investigated. This class of algorithms enables decoding at extremely-high data rates (up to some hundred Gbps), but it comes at the cost of sacrificing some coding gain, especially at moderate-low code rates.
On the hardware side, pipelined \ac{LDPC} decoder architectures promise to achieve unmatched decoding speeds by ``unrolling'' the \ac{BP} decoder iterations over the chip, hence realizing a fully parallel decoder without the message routing hurdle that affects non-pipelined decoder architectures \cite{Wehn13}. Remarkably, pipelined \ac{LDPC} decoder architectures support soft-decision \ac{BP} decoding at data rates exceeding $100$ Gbps.\footnote{A pipelined \ac{LDPC} decoder supporting data rates up $160$ Gbps was demonstrated in \cite{Wehn13}. The data rate refers to a $65$nm ASIC design with a clock frequency of $257$ MHz, with a $4$-bits messages quantization.}
While decoding algorithms and architectures are obviously impacted by the need to operate at large data rates, other elements of the communication chain can be affected, too. In wireless systems operating in high frequency bands (e.g, in W/D bands, as currently considered for $6$G cellular networks), low-order modulation schemes, such as \ac{OOK}, \ac{BPSK}, and \ac{QPSK}, are considered mainly thanks to their simplicity and to the abundant bandwidth available  \cite{Alo21}. Regular framing structures, obtained by inserting periodic \ac{SoF} markers, are also preferred since they simplify the signal acquisition \cite{Morello}. These observations place further constraints on channel code design, which can be summarized by (a) the need of constructing a \emph{single} code (to support pipeline decoder architectures), (b) with rate flexibility that should be achieved \emph{without} modifying the blocklength (to enable periodic \ac{SoF} markers). Focusing the attention to the class of \ac{LDPC} codes, (a) implies the implementation of a pipeline decoder that unrolls the Tanner graph of a single \ac{LDPC} code, while (b) rules out classical rate adaptation techniques such as puncturing \cite{Yaz04,McL06}, shortening, and lengthening \cite{Ngu12,Divsalar14:ProtoRaptor,Rag19} that are currently employed by the $3$GPP $5$G-NR standard \cite{Richardson5G}.

A class of multi-rate \ac{LDPC} codes that facilitate the adoption of a unified decoder architecture while keeping a constant blocklength was introduced in \cite{Wesel09}. The approach of \cite{Wesel09} relies on expurgation of an high-rate \ac{LDPC} code, constructing \ac{LDPC} codes of lower rate. In particular, the construction of \cite{Wesel09} is based on the following observation: starting from a low-rate \ac{LDPC} code (referred to as \emph{mother code} in \cite{Wesel09}), it is possible to obtain higher-rate \ac{LDPC} codes by linearly combining rows of the mother code parity-check matrix. The ingenuity of the approach of \cite{Wesel09} stems from the construction of the low-rate code, and on the choice of rows to be linearly combined, targeting a common decoder architecture and efficient (i.e., linear-time) encoding of all code rates. In particular, ancillary degree-$2$ check nodes are introduced in the mother code Tanner graph to reflect the combining of parity-check equations. By selectively activating the ancillary degree-$2$ check nodes, a single decoder architecture can be used to decode all codes derived from the mother code. While this approach is simple and elegant, it is not clear what challenges it entails when pipeline decoder architectures are considered. Furthermore, additional engineering of the row combining approach is required to efficiently support low code rates \cite{Wesel09}.

\ac{MN} codes  were introduced in \cite[Sec. VI]{Mac99} as a class of error correcting codes based on sparse matrices for nonuniform sources. 
\ac{MN} codes are multi-edge-type  \ac{LDPC} codes \cite{RU08}. The Tanner graph of a \ac{MN} code can be split in two parts, with a set of \acp{VN} associated with the source bits, and the remaining \acp{VN} associated with the codeword bits. \ac{LDPC} code constructions closely related to \ac{MN} codes were proposed in \cite{Wainwright,fresia2010joint} for joint source and channel coding. While originally introduced to deal with nonuniform sources, it was pointed out in \cite{Mac99} that \ac{MN} can also be employed with uniform sources by introducing a nonlinear block code that turns the uniform source output sequence into a sequence with a prescribed distribution. The potential appeal of this construction, as observed in \cite{Mac99}, stems from the possibility of modifying the code rate (e.g., adapting it to varying channel conditions) by changing the statistics of the sequences produced by the nonlinear block encoder, hence without modifying the underlying \ac{LDPC} code. With reference to the constraints elaborated in the previous paragraphs, an important consequence is that a rate-adaptive scheme based on \ac{MN} codes can keep the blocklength constant. Furthermore, since decoding is performed over a fixed Tanner graph---regardless of the code rate defined by the outer nonlinear encoder---\ac{MN} codes are well suited for a pipeline decoder architecture.

From a theoretical viewpoint, some attention was placed in showing the capacity-achieving properties of \ac{SC} \ac{MN} code ensembles in \cite{Kasai11,Mit12,Oba13}. However, \ac{MN} codes as a means to achieve rate flexibility received little attention. A notable exception is the \ac{PAS} scheme introduced in \cite{bocherer2015bandwidth}, where a construction reminiscent of \ac{MN} codes was proposed. In \ac{PAS}, the sequence output by a uniform (binary) source is also processed by the nonlinear block encoder, referred to as \ac{DM} \cite{bocherer2015bandwidth}, generating a sequence of amplitude symbols with a given empirical distribution. The binary labels associated with the amplitude symbols are encoded through a nonsystematic \ac{LDPC} code encoder, producing a parity bit vector. The amplitude symbols together with the parity bits are then mapped onto \ac{PAM} symbols. The rates achievable by \ac{PAS} were analyzed in \cite{Boecherer2017b,Amjad2018}  showing that the layered shaping architecture consisting of \ac{PAS}, together with a decoder that performs a search over the entire inner codebook, is capacity-achieving under a maximum a posteriori probability decoding metric (see also \cite{bocherer2023probabilistic,Merhav:2023a}).
While the main result attained by \ac{PAS} is to provide sizable shaping gains, it was quickly recognized that, as a byproduct, \ac{PAS} is naturally rate-adaptive thanks to the possibility of tuning the \ac{DM} rate \cite{bocherer2015bandwidth,Buchali2016}, as originally hypothesized in \cite{Mac99}. Differently from the approach outlined in \cite{Mac99}, \ac{PAS} targets high spectral efficiencies by shaping the probability of the amplitudes of the constellation symbols, and cannot be readily used to adapt the rate in binary modulation schemes. In the context of \ac{PAS}, several classes of \acp{DM} have been proposed and analyzed \cite{Schulte2016,Schulte19,Schu20}, providing an extensive understanding on how to design efficient \ac{DM} algorithms for high-speed communications.

In this paper, we analyze rate-adaptive \ac{MN} codes based on protographs \cite{Tho03}.  The analysis is provided for the \ac{biAWGN} channel and as \ac{DM} we consider \acp{CCDM} \cite{bocherer2015bandwidth,Schulte2016}. The extension of the analysis to general memoryless binary-input output-symmetric channels and to the use of other classes of \acp{DM} is trivial. Noting that the concatenation of the \ac{CCDM} with the inner linear block code results in a nonlinear code, we introduce an equivalent communication model that simplifies the analysis of both \ac{ML} and \ac{BP} decoders. In particular, the equivalent model resorts to the study of the performance of the protograph \ac{LDPC} code over the communication channel, where side information is provided to the decoder by observing the \ac{LDPC} encoder input through a binary-input, binary-output symmetric channel. Leveraging on this construction, we analyze protograph \ac{MN} code ensembles in terms of iterative decoding threshold, both via \ac{PEXIT} analysis \cite{ten01,Liva2007:EXIT_GC} and via a more accurate quantized \ac{DE} analysis \cite{RU01a}. Via numerical examples, we show that while the \ac{PEXIT} analysis provides a fast and accurate estimate of the \ac{BP} decoding thresholds at medium-high code rates, it tends to produce unreliable estimates at very low code rates. We hence propose a  protograph \ac{MN} code ensemble design via the (faster) \ac{PEXIT} analysis, supplemented by the more accurate quantized \ac{DE} analysis at low rates. The distance properties of protograph \ac{MN} code ensembles are studied, showing how the input-output weight enumerator of the inner \ac{LDPC} codes can be used to analyze the error floor performance. By means of asymptotic enumeration techniques, we introduce a criterion on the shape of the normalized logarithmic asymptotic input-output weight distribution that allows discarding code ensembles that are likely to yield codes with high error floors. Numerical results for selected code design examples confirm the accuracy of the analysis, which enables the design of \ac{MN} codes with thresholds within approx. $1$ dB from the Shannon limit, over a wide range of code rates (where each code rate is obtained by tuning the \ac{DM} parameters, without modifying the inner \ac{LDPC} code).

The paper is organized as follows. Section \ref{sec:prel} provides preliminary definitions. Protograph \ac{MN} codes are described in Section \ref{sec:MNcodes}, whereas their decoding is discussed in Section \ref{sec:decoding}. Section \ref{sec:decoding} includes also the definition of the equivalent communication model that enables the analysis of the decoder performance. The \ac{DE} analysis is provided in Section \ref{sec:DE}, whereas the analysis of the distance properties of protograph \ac{MN} code ensembles is developed in Section \ref{sec:WEF}, together with the derivation of a union bound on the error probability of \ac{MN} codes. Some detailed steps in the distance spectrum analysis are contained in Appendix \ref{appendix:IOWEF}. The code design methodology is outlined in Section \ref{sec:design}. Numerical results and conclusions follow in Section \ref{sec:results} and in Section \ref{sec:conclusions}, respectively.

%%%%%%%%%%%%%%%%%%%%%%%%%%%%%%%%%%%%%%%%%%%%%%%%%%%%%%%%%%%%%%%%%%%%%%%%%
%%%%%%%%%%%%%%%%%%%%%%%%%%%%%%%%%%%%%%%%%%%%%%%%%%%%%%%%%%%%%%%%%%%%%%%%%
%%%%%%%%%%%%%%%%%%%%%%%%%%%%%%%%%%%%%%%%%%%%%%%%%%%%%%%%%%%%%%%%%%%%%%%%%

\section{Preliminaries}\label{sec:prel}

We denote \acp{r.v.} by uppercase letters and their realizations by lowercase letters. The \ac{p.m.f.} of a discrete \ac{r.v.} $X$ is $P_X(x)= \prob{X=x}$, and the \ac{p.d.f.} of a continuous \ac{r.v.} $X$ is $p_X(x)$. In either case, the subscript will be dropped whenever no ambiguity may arise, i.e., $P(x)=P_X(x)$ and $p(x)=p_X(x)$. We use $\Hb(p)=-p \log_2 p -(1-p)\log_2(1-p)$, with $0<p<1$ and $\Hb(0)=\Hb(1)=0$, for the binary entropy function. Similarly, $H(p)=-p\ln(p)-(1-p)\ln(1-p)$ denotes the natural binary entropy function. Row vectors are denoted by bold letters, e.g., $\bm{x}$, while matrices are denoted by uppercase bold letters, e.g., $\vX$. The order-$2$ finite (Galois) field is $\fieldtwo$. Finally, $\hw(\bm{x})$ and $\hd(\bm{x}, \bm{y})$ are the Hamming weight of a vector $\bm{x}$ and the Hamming distance between two vectors $\bm{x}$ and $\bm{y}$, respectively. 

\subsection{Channel Model}\label{sec:prel:AWGN}
We consider transmission of a \ac{BPSK} modulated signal over the \ac{AWGN} channel. The resulting memoryless discrete-time \ac{biAWGN} channel model is defined by $Y = X + N$ where $Y$ is the channel output, $X\in\{-1,+1\}$ is the channel input, and where $N\sim\mathcal{N}(0,\sigma^2)$ is the additive white Gaussian noise term. The channel \ac{SNR} is defined as $E_s/N_0 = 1/(2\sigma^2)$ where $E_s$ is the energy per \ac{BPSK} symbol and $N_0$ is the single-sided noise power spectral density.

\subsection{Protograph LDPC Codes}\label{sec:prel:proto}

A protograph $\pgraph = (\cV , \cC, \edges)$ is a small Tanner graph \cite{Tan81}  consisting of a set $\cV$ of  $\mathtt{N}$ \acp{VN}, a set $\cC$ of $\mathtt{M}$ \acp{CN}, and a set $\edges$ of $e$ edges \cite{Tho03}. \acp{VN} in the protograph are numbered from $1$ to $\mathtt{N}$. Similarly, protograph \acp{CN} are numbered from $1$ to $\mathtt{M}$. Each \ac{VN}/\ac{CN}/edge in a protograph defines a \ac{VN}/\ac{CN}/edge type.  We denote by $\edges_{\vn_j}$ ($\edges_{\cn_i}$)  the set of edges in the protograph connected to $\vn_j$ ($\cn_i$).  The degree $\dvj$ of $\vn_j$ ($\dci$ of $\cn_i$) is then equal to $|\edges_{\vn_j}|$ ($|\edges_{\cn_i}|$). The Tanner graph $\bgraph$ of an \ac{LDPC} code can be derived by lifting the protograph. In particular, the protograph is copied ${\liftfac}$ times (where ${\liftfac}$ is referred to as the \emph{lifting factor}), and the edges of the protograph copies are permuted under the following constraint: if an edge connects a type-$j$ \ac{VN} to a type-$i$ \ac{CN} in $\pgraph$, after permutation the edge should connect one of the ${\liftfac}$ type-$j$ \ac{VN} copies with one of the ${\liftfac}$ type-$i$ \ac{CN} copies in $\bgraph$. We denote by $\vn_1,\vn_2,\ldots$  \acp{VN} in $\bgraph$, and by $\cn_1,\cn_2,\ldots$ the \acp{CN} in $\bgraph$.
The lifted graph $\bgraph$ defines the parity-check matrix of an \ac{LDPC} code.  The base matrix of a protograph is an $\mathtt{M} \times \mathtt{N}$ matrix $\vB= [b_{i,j}]$ where $b_{i,j}$ is the number of edges that connect \ac{VN} $j$ to \ac{CN} $i$ in $\pgraph$. We will make use of \ac{LDPC} codes with \emph{punctured} (or \emph{state}) \acp{VN}. A punctured \ac{VN} is associated with a codeword bit that is not transmitted through the communication channel. We will assume that all the \acp{VN} of a given type are either punctured or they are not, i.e., puncturing is already defined at protograph level.

\begin{figure}
	\begin{center}
		\includegraphics[width=0.6\columnwidth]{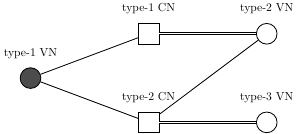}
		\caption{Protograph of Example \ref{ex:proto}.}\label{fig:protoexample}
	\end{center}
\end{figure}

\begin{figure}
	\begin{center}
		\includegraphics[width=0.9\columnwidth]{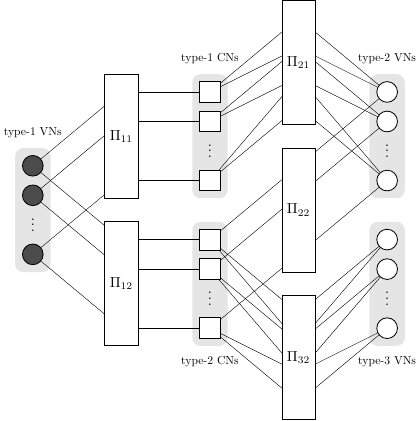}
		\caption{Tanner graph obtained by lifting the protograph of Example \ref{ex:proto}.}\label{fig:protoexp}
	\end{center}
\end{figure}

	\begin{example}\label{ex:proto}
		Consider the $2\times 3$ protograph base matrix
		\[
		\vB = \left(
		\begin{array}{c|cc}
			1 &  2 & 0\\
			1 &  1 & 2\\
		\end{array}
		\right).
		\]
		The vertical line partitions the base matrix in two parts: a $2\times1$ left submatrix, and a $2\times 2$ right submatrix. The left submatrix is associated with punctured \acp{VN}. The corresponding protograph is shown in Figure~\ref{fig:protoexample}. The type-$1$ protograph \ac{VN} is represented by a dark circle to emphasize that the \acp{VN} of type $1$ are punctured. The Tanner graph obtained by lifting the protograph is depicted in Figure~\ref{fig:protoexp}. The edge lifting is described by means of the edge interleavers $\Pi_{ij}$, where $\Pi_{ij}$ defines the permutation applied to the edges connecting type-$i$ \acp{VN} to type-$j$ \acp{CN}. 
	\end{example}

A protograph (base matrix) defines a protograph \ac{LDPC} code ensemble. More specifically, given a protograph $\pgraph$ and a lifting factor $\liftfac$, the ensemble is given by the codes whose graph $\bgraph$ can be obtained by an $\liftfac$-fold protograph lifting. With reference to Example \ref{ex:proto}, a random code from the ensemble can be obtained by drawing each edge interleaver uniformly at random within the set of all possible permutations.

\subsection{Asymptotic Enumeration Results}
Let $ a(n) $ and $ b(n) $ be two real-valued sequences, where $b(n) \neq 0 \ \forall n $. Then, $ a(n) $ is \emph{exponentially equivalent} to $ b(n) $ as $ n \rightarrow \infty $ if and only if \cite[Sec. 3.3]{Cover2006} 
\[
\lim\limits_{n \rightarrow \infty} \frac{1}{n} \ln\frac{a(n)}{b(n)} =0.
\]
We will use the notation $ a(n) \dot= b(n) $ to specify that  $a(n) $ is exponentially equivalent to $ b(n) $. Moreover, given $\bm{z} =( z_{1}, z_{2},\ldots, z_{d})  $ and $ \bm{\beta} = ( \beta_{1}, \beta_{2}, \ldots, \beta_{d}) $, we {use the shorthand} 
\begin{equation*}
	\bm{z}^{\bm{\beta}} = \prod\limits_{t=1}^{d} z_t ^{\beta_t}.
\end{equation*}
In the distance spectrum analysis of \ac{MN} codes, we will leverage efficient asymptotic enumeration methods. In particular, we will make use of the following result.
\begin{lemma} \label{lemma:lemmaHayman} [Hayman Formula for Multivariate Polynomials \cite[Corollary 16]{DUR06}
	Let $ \bm{z} = ( z_{1}, z_{2},\ldots, z_{d}) $ and let $ p (\bm{z}) $ be a multivariate polynomial with $ p(\bm{0}) \neq 0 $. Let $ \bm{\beta} = ( \beta_{1}, \beta_{2}, \ldots, \beta_{d}) $ where $ 0 \leq \beta_{t} \leq 1 $ and $ \beta_{t}n $ is an integer for all $ t \in \left\lbrace 1, 2, \ldots, d\right\rbrace $. Then \begin{equation*}
		\coeff\left( ( p(\bm{z})) ^{n} , \bm{z} ^{n\bm{\beta}} \right)  \dot = \exp\left\{n \left[ \ln p(\bm{z}^\ast)  - \sum\limits_{t=1}^{d} \beta_{t} \ln z^\ast_{t} \right]\right\}
	\end{equation*}
	where $  \coeff\left(  p(\bm{z}) ^{n} , \bm{z} ^{n \bm{\beta}} \right) $ represents the coefficient of $\bm{z} ^{n\bm{\beta}} $ in the polynomial $p(\bm{z}) ^{n}$,  $ \bm{z}^\ast = ( z^\ast_{1}, z^\ast_{2},\ldots, z^\ast_{d}) $ and $ z^\ast_{1}, z^\ast_{2},\ldots, z^\ast_{d} $ are the unique positive solutions to
	\begin{equation*}
		z_{t} \dfrac{\partial p (\bm{z}) }{ \partial z_{t}} = \beta_{t}  p (\bm{z}), \hspace{1cm} \forall t  \in \left\lbrace 1, 2, \ldots, d\right\rbrace. 
	\end{equation*}
\end{lemma}

%%%%%%%%%%%%%%%%%%%%%%%%%%%%%%%%%%%%%%%%%%%%%%%%%%%%%%%%%%%%%%%%%%%%%%%%%
%%%%%%%%%%%%%%%%%%%%%%%%%%%%%%%%%%%%%%%%%%%%%%%%%%%%%%%%%%%%%%%%%%%%%%%%%
%%%%%%%%%%%%%%%%%%%%%%%%%%%%%%%%%%%%%%%%%%%%%%%%%%%%%%%%%%%%%%%%%%%%%%%%%

\section{Protograph MacKay-Neal Codes}\label{sec:MNcodes}

\begin{figure*}
	\begin{center}
		\includegraphics[width=16cm]{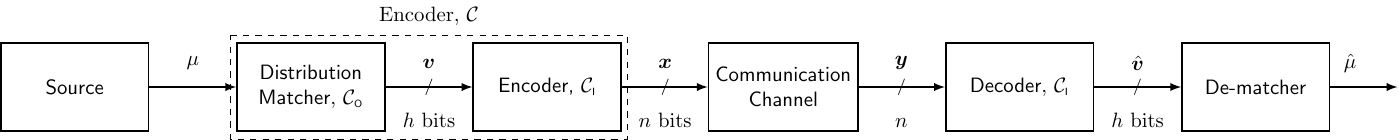}
		\caption{System model, where a \ac{MN} code is used to communicate over the \ac{biAWGN} channel (communication channel).}\label{fig:model}
	\end{center}
\end{figure*}

\ac{MN} codes were originally introduced as a class of \ac{LDPC} codes with nonsystematic encoding to be used with nonuniform sources \cite[Sec. VI]{Mac99}. It was suggested that \ac{MN} codes can be used for uniform sources, too, by introducing an outer nonlinear code (e.g., obtained by reversing the role of the encoder and the decoding in a standard arithmetic codec) in concatenation with the inner nonsystematic \ac{LDPC} encoder \cite[Sec. VI.A]{Mac99}. Moreover, it was recognized that the latter construction can be used to adapt the code rate when communicating over channels with different noise levels without modifying the Tanner graph of the inner \ac{LDPC} code \cite[Sec. VI.C]{Mac99}. In the following, we refer to \ac{MN} codes in this second flavor, i.e., as a class of codes obtained by concatenating an outer nonlinear code $\outcode$ with an inner linear block code $\incode$. More specifically, we consider the setting depicted in Figure \ref{fig:model}. A uniform source generates a message $\mu \in \{1,2,\ldots,M\}$. The message is input to the encoder of a length-$h$ outer code $\outcode$. The outer encoder generates an output sequence with a prescribed empirical distribution, i.e., $\outcode$ is a \ac{CC} code. Thus, each codeword of $\outcode$ has the same Hamming weight. We refer to the \ac{CC} encoder as the \emph{distribution matcher} (DM) \cite{bocherer2015bandwidth}. Note that, by restricting our attention to outer \ac{CC} codes, we can leverage low-complexity outer code encoders based on arithmetic coding techniques \cite{Mac99,Schulte2016}. Let $\omega\in(0,1)$ denote the fractional Hamming weight of the $h$-bits outer \ac{CC} codeword $\vecv$, i.e., $\omega = \hw(\vecv)/h$. We have that 
$M = |\outcode|={h \choose \omega h}$. Hence, the rate of the outer code is 
\[
\dmR = \frac{1}{h}\log_2 M = \frac{1}{h}\log_2 {h \choose \omega h}
\]
which converges to $\Hb(\omega)$ for large $h$. The output of the \ac{DM} is then input to the encoder of an inner $(n,h)$ binary linear block code $\incode$ defined by
\begin{equation}
	\incode = \left\{ \vecc \big| \vecc \vH_2^\transposed = \vecv \vH_1^\transposed, \vecv \in \fieldtwo^h\right\}
\end{equation}
where $\vH_1$ is an $n \times h$ sparse binary matrix, and $\vH_2$ is an $n \times n$ sparse invertible binary matrix. Note that, strictly speaking, $\incode$ may not be an \ac{LDPC} code, i.e., the code may not possess a sparse parity-check matrix. Nevertheless, $\incode$ can be seen as the code obtained by puncturing an $(n+h,h)$ \ac{LDPC} code with $n\times (h+n)$ parity-check matrix 
\begin{equation}
	\vH = \left[\,\vH_1\, |\, \vH_2\,\right] \label{eq:LDPCH}
\end{equation}
where puncturing is applied to the first $h$ coordinates. We refer to the $(n+h,h)$ \ac{LDPC} code with parity-check matrix in the form \eqref{eq:LDPCH} as the (inner) \emph{mother code} $\inmcode$. 
The inner code rate is $\inR = h/n$, whereas the mother code rate is $\inmR = h / (n+h) = \inR / (1 + \inR)$. The nonsystematic generator matrix of the inner code $\incode$ is $\bm{G}=\vH_1 ^{\transposed}\vH_2^{-\transposed}$. By observing that the inverse of $\vH_2$ is generally dense, we have that $\bm{G}$ is dense too.
Hence, the code generated by the concatenation of $\outcode$ with $\incode$ results in a marginal distribution of the codeword bits that is very close to the uniform one \cite{bocherer2015bandwidth}. This result is fundamental since the capacity-achieving input distribution of the \ac{biAWGN} channel is uniform.

We denote by $\code$ the overall code resulting from the concatenation of the inner and outer codes. The rate of $\code$ is 
\begin{equation}
	R = \dmR \inR \approx \Hb(\omega) \inR.   \label{eq:rates} 
\end{equation}
As observed in \cite[Sec. VI.C]{Mac99}, all rates $0<R<\inR$ can be achieved simply by fixing the \ac{DM} parameter $\omega$, without requiring any modification (e.g., puncturing/shortening) of the inner code. An \ac{MN} code is fully defined by the parity-check matrix of the inner mother code $\inmcode$ and by the \ac{DM} parameter $\omega$. We refer to a \ac{MN} code \emph{family} as the set of \ac{MN} codes with fixed mother code, obtained for all  $\omega \in (0,1)$ that yield an integer $\omega h$.

\subsection{Protograph-based Construction}\label{sec:MNcodes:Proto}

\ac{MN} can be constructed by protograph expansion. In particular, a protograph \ac{MN} code is obtained by concatenating the outer \ac{CC} code with a protograph mother \ac{LDPC} code $\inmcode$: the mother code parity-check matrix \eqref{eq:LDPCH} is obtained by lifting a protograph whose $n_0 \times (h_0 + n_0)$ base matrix takes the form
$\vB = \left[\,\vB_1\, |\, \vB_2\,\right]$
where $\vB_1$ is $n_0 \times h_0$ and $\vB_2$ is $n_0 \times n_0$, with integers $n_0=n/\liftfac$ and $h_0=h/\liftfac$. All type-$i$ \acp{VN} with $i=1,\ldots,h_0$ are punctured. A (protograph) \ac{MN} code ensemble  $ \ensP{\pgraph}{\omega} $ is the set of \ac{MN} codes whose inner mother code Tanner graph $\bgraph$ is obtained by lifting $\pgraph$, and where the \ac{DM} parameter is $\omega$. The \ac{MN} code ensemble family $\ensfP{\pgraph}$ is the set of ensembles  $\left\{\ensP{\pgraph}{\omega}\right\}_{\omega \in (0,1)}$.

\section{Decoding of MN Codes}\label{sec:decoding}

With reference to the \ac{biAWGN} channel model (Section \ref{sec:prel:AWGN}), the codeword $\vecc$ is mapped onto $\{-1,+1\}^n$ via binary antipodal modulation through $x_i = 1-2c_i$, for $i=1,\ldots,n$. 
With a slight abuse of wording, we will refer to the modulated codeword $\vecx$ as the codeword. Similarly, we will use $\incode$ and $\code$ to denote the modulated codebook of the inner code and of the overall code, respectively. 

In the following, we first review the \ac{BP} decoding algorithm applied to \ac{MN} codes (Section \ref{sec:decoding:BPdecoding}). We provide then  a discussion of general decoding metrics (Section \ref{sec:decoding:decoding_metrics}), that will prove useful to analyze the performance of \ac{MN} codes thanks to their interpretation in the context of an equivalent parallel channel model (Section \ref{sec:MNcodes:equivmodels}).

\subsection{Belief Propagation Decoding}\label{sec:decoding:BPdecoding}

\begin{figure}
	\centering
		\includegraphics[width=\columnwidth]{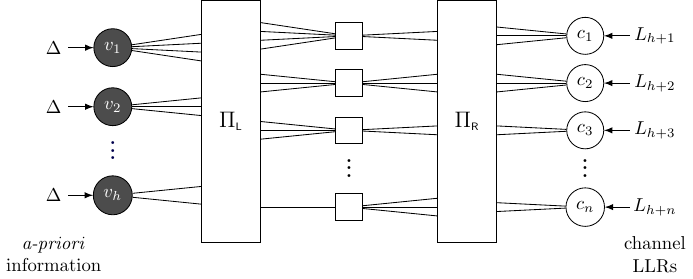}
	\caption{Belief propagation decoding over the Tanner graph of the mother \ac{LDPC} code.}
	\label{fig:BP}
\end{figure}

\ac{MN} codes can be conveniently decoded via the \ac{BP} algorithm over the Tanner graph of the inner mother \ac{LDPC} code. 
Let us denote by $L_i$ the $L$-value at the input of the $i$th \ac{VN}. 
The \ac{BP} decoder is initialized by setting
\begin{equation}
	L_i = \left\{
	\begin{array}{lll}
		\Delta  &  &\qquad 1\leq i\leq h\\
		\ln \displaystyle\frac{p(y_{i-h}|0)}{p(y_{i-h}|1)} &\qquad &\qquad h<i\leq h+n.
	\end{array}
	\right.
\end{equation}  
Here, 
\[
\Delta := \ln \frac{1-\omega}{\omega}
\]
while $p(y_i|c_i)$ 
is the probability density of the \ac{biAWGN} channel output $y_i$ conditioned on the transmission of the codeword bit $c_i$. Hence, $L_i=2y_{i-h}/\sigma^2$ for $h<i\leq h+n$. The initialization of the \ac{BP} decoder is displayed in Figure \ref{fig:BP}, where the Tanner graph of the inner mother \ac{LDPC} code is shown. In the figure, $\Pi_{\scaleto{\mathsf{L}}{4pt}}$ and $\Pi_{\scaleto{\mathsf{R}}{4pt}}$ represent the edge permutations associated with the submatrices $\bm H_1$ and $\bm H_2$ in \eqref{eq:LDPCH}. The punctured \acp{VN} associated to the bits $v_1, v_2, \ldots, v_h$ are represented as dark circles. Observe that the punctured \acp{VN} are provided with prior information resulting from the marginal distribution of the \ac{CC} code codeword $\vecv$. For \acp{VN} associated with the codeword bits that are transmitted through the \ac{biAWGN} channel, we input the corresponding channel \acp{LLR}.

\ac{BP} decoding proceeds through the standard, iterative message-passing algorithm over the code graph. After reaching a fixed maximum number of iterations, the decoder outputs the decision
$\hat{\vecv}$. If the composition (i.e., Hamming weight) of $\hat{\vecv}$ differs from the one defined by the outer \ac{CC} code, an error is declared. Otherwise,   $\hat{\vecv}$ is processed by the de-matcher \cite{Schulte2016}, producing the estimate $\hat{\mu}$ of the transmitted message (see Figure \ref{fig:model}).

Note that the decoder outlined above (proposed already in \cite{Mac99}) employs the same layered decoding architecture adopted by \ac{PAS} schemes \cite{bocherer2015bandwidth,Boecherer2017b}: the \ac{BP} decoder does not have any information on the outer \ac{CC} constraints, and it exploits only the knowledge of the marginal distribution of the bits in $\vecv$.

\subsection{Maximum-likelihood and Mismatched Decoding}\label{sec:decoding:decoding_metrics}

We discuss \ac{ML} decoding of \ac{MN} codes, which will be useful to analyze the code performance in the error floor region. Given an \ac{MN} code $\code$, the \ac{ML} decoder outputs 
\begin{equation}
	\xML = \argmax_{\vecx\in \code} p(\vecy|\vecx) \label{eq:ML1}
\end{equation}
where $p(\vecy|\vecx)$ is the probability density of the \ac{biAWGN} channel output $\vecy$ conditioned on the input $\vecx$. Note that the \ac{ML} criterion \eqref{eq:ML1} can be rephrased as
\begin{equation}
	\xML = \argmax_{\vecx\in \incode} p(\vecy|\vecx)P(\vecv[\vecx]). \label{eq:ML2}
\end{equation}
In \eqref{eq:ML2}, the bijective relation between $\vecx$ (inner encoder output) and $\vecv$ (inner encoder input) is emphasized by introducing the notation $\vecv[\vecx]$. Note that $P(\vecv)$ takes value $1/|\outcode|$ if $\vecv \in \outcode$, whereas $P(\vecv)$ equals zero if 
$\vecv \notin \outcode$. Furthermore, the decoder defined by \eqref{eq:ML2} performs a search over the inner code $\incode$, hence, over an enlarged set compared to \eqref{eq:ML1}: the outer code constraints are conveyed by the prior $P(\vecv)$. 

It is fundamental to observe that the \ac{BP} decoder outlined in Section \ref{sec:decoding:BPdecoding} cannot exploit the joint distribution of the bits forming $\vecv$,\footnote{\ac{BP} decoding can be extended to exploit the joint distribution of the bits in $\vecv$ by iterating between the inner mother \ac{LDPC} decoder and an outer soft-input soft-output decoder designed for the \ac{CC} code. The outer \ac{CC} decoder can be based, for instance, on the forward-backward algorithm applied to the trellis representation of the \ac{CC} code. As shown in \cite{schulte2020joint}, the technique allows to improve the performance of \ac{PAS} schemes when moderate-small blocklengths are considered. The gains are negligible for large blocklenghts.} but it rather uses the marginal distribution of the bits composing $\vecv$ to bias the decoder operating over the Tanner graph of the mother code. It is hence of interest to analyze the \emph{mismatched} decoding rule
\begin{equation}
	\xMM = \argmax_{\vecx\in \incode} p(\vecy|\vecx)Q(\vecv[\vecx]). \label{eq:MM1}
\end{equation}
where $Q(\vecv)=\prod_{i=1}^h P(v_i)$ denotes the product of the marginal distributions of the bits $v_1, v_2,\ldots,v_h$. 
The decoding metric adopted in \eqref{eq:MM1} is suboptimal compared with the one of \eqref{eq:ML2}. In fact, the term $Q(\vecv)$ acts as a mismatched prior, yielding a nonzero probability also for $\vecv \notin \outcode$. 

\subsection{Equivalent Parallel Channel Model}\label{sec:MNcodes:equivmodels}

Analyzing the performance of \ac{MN} codes is challenging, even under the decoding metrics of the previous subsection and over \ac{MBIOS} channels. This fact is mostly due to the nonlinear nature of $\code$, which renders the block error probability under \eqref{eq:ML1}--\eqref{eq:ML2}  dependent on the transmitted codeword, hindering the use of a reference codeword to compute bounds on the block error probability. The same issue arises when attempting a \ac{DE} evolution analysis under \ac{BP} decoding, where the {allzero} codeword is often used as a reference.

\begin{figure*}
	\centering
	{
		\includegraphics[width=16cm]{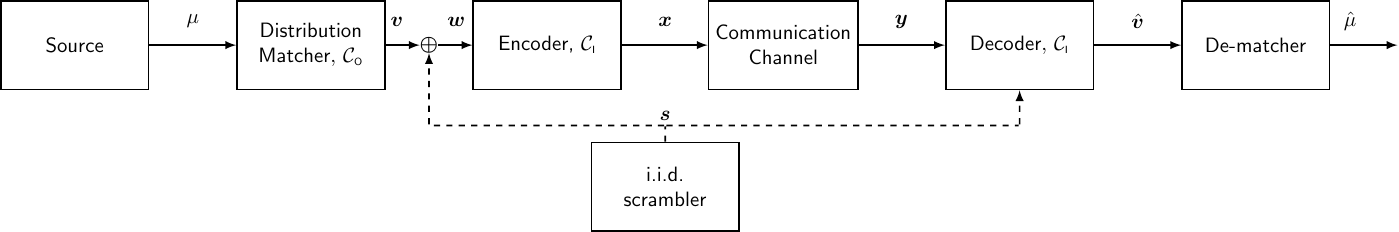}
	}
	\caption{Modification of the system model of Figure \ref{fig:model}, where an i.i.d. scrambler is introduced.}
	\label{fig:equiv1}
\end{figure*}

The issue can be circumvented by resorting to alternative communication models \cite{wyner1974recent,fresia2010joint} that can be proved to be equivalent to the one depicted in Figure \ref{fig:model}. Consider first the scheme depicted in Figure \ref{fig:equiv1}, where a scrambling block is introduced. The block generates a sequence $\vecscr=(\sscr_1,\sscr_2,\ldots,\sscr_h)$ where each element is picked independently and uniformly at random in $\{0,1\}$. The sequence $\vecscr$ is then added (in $\fieldtwo$) to $\vecv$. The sum of the two vectors, denoted by $\vecw$, is then encoded with $\incode$. Note that  $\vecv$ and $\vecscr$ are independent and that the marginal distributions of the entries of $\vecv$ and $\vecscr$ are Bernoulli with parameters $\omega$ and $1/2$, respectively. It follows that the entries in $\vecw$ are uniformly distributed, i.e., they follow a Bernoulli distribution with parameter $1/2$. The sequence $\vecscr$ is assumed to be available to the decoder. Considering either \eqref{eq:ML1} or \eqref{eq:ML2}, and owing to the symmetry of the \ac{biAWGN} channel, we observe that the presence of the scrambler is irrelevant to the analysis of the error probability, since the addition of $\vecscr$ at the transmitter side can be compensated at the decoder by computing first $\vecb=\vecscr \bm{G}$, and then flipping the sign of the observations $y_i$ for all $i\in \mathrm{supp}(\vecb)$.

\begin{figure}
	\centering
	{
		\includegraphics[width=8cm]{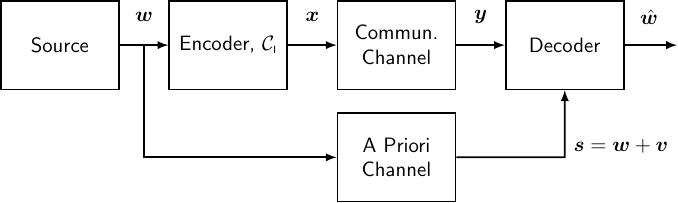}
	}
	\caption{Equivalent parallel channel model.}
	\label{fig:equiv2}
\end{figure}

The analysis of the bit/block error probability under the model of Figure \ref{fig:equiv1}, averaged over all possible transmitted codewords, is equivalent to the analysis of the bit/block error probability of the communication model described in Figure \ref{fig:equiv2}: here, an \ac{i.i.d.} uniform binary source generates an $h$-bits vector $\vecw$, which is encoded via $\incode$ yielding a codeword $\vecx$ that is transmitted through the \ac{biAWGN} channel. The decoder also obtains an observation of $\vecw$ via a so-called \emph{a priori channel}. The a priori channel adds (in $\fieldtwo$) a weight-$\omega h$ binary vector $\vecv$ to $\vecw$, were $\vecv$ is picked uniformly at random in $\outcode$, resulting in the observation $\vecscr$. Upon observing $\vecy$ and $\vecscr$, the decoder produces a decision on $\vecw$, or, equivalently, a decision on $\vecv$ since $\vecw=\vecv+\vecscr$. \ac{ML} decoding will produce
\begin{equation}
	\xML = \argmax_{\vecx\in \incode} p(\vecy|\vecx)P(\vecscr|\vecw[\vecx]) \label{eq:ML3}
\end{equation}
where $P(\vecscr|\vecw)=1/|\outcode|$ if $(\vecscr-\vecw) \in \outcode$, and $P(\vecscr|\vecw)=0$ otherwise. The decoding problem can be immediately recognized to be equivalent to the one in \eqref{eq:ML2}. The decoder may also resort to a mismatched model for the a priori channel, treating it as a \ac{BSC} with crossover probability $\omega$ and resulting in 
\begin{equation}
	\xMM = \argmax_{\vecx\in \incode} p(\vecy|\vecx)Q(\vecscr|\vecw[\vecx]) \label{eq:ML4}
\end{equation}
where $Q(\vecscr|\vecw)=\prod_{i=1}^h P(z_i|w_i)$, i.e., the solution of \eqref{eq:ML4} is equivalent to the solution of \eqref{eq:MM1}.
We refer to the model of Figure \ref{fig:equiv2} as the \ac{EPC} \emph{model}. 

\begin{remark}The convenience of the \ac{EPC} model stems from the fact that, owing to the symmetry of the communication and of the a priori channels and to the linearity of the code $\incode$, the error probability is independent on $\vecw$: we can analyze the error probability of the scheme of Figure \ref{fig:equiv2} by fixing as reference the allzero codeword. The resulting analysis will characterize exactly the error probability of the original scheme of Figure \ref{fig:model}, averaged over all possible transmitted codewords.
\end{remark}

%%%%%%%%%%%%%%%%%%%%%%%%%%%%%%%%%%%%%%%%%%%%%%%%%%%%%%%%%%%%%%%%%%%%%%%%%
%%%%%%%%%%%%%%%%%%%%%%%%%%%%%%%%%%%%%%%%%%%%%%%%%%%%%%%%%%%%%%%%%%%%%%%%%
%%%%%%%%%%%%%%%%%%%%%%%%%%%%%%%%%%%%%%%%%%%%%%%%%%%%%%%%%%%%%%%%%%%%%%%%%

\section{Density Evolution Analysis}\label{sec:DE}

The \ac{EPC} model introduced in Section \ref{sec:MNcodes:equivmodels} allows to analyze \ac{MN} code ensembles from a \ac{DE} viewpoint. The analysis follows by computing the \ac{DE} recursions for the mother  \ac{LDPC} code protograph, where the initialization of the message densities accounts for the type of channel associated to the protograph \acp{VN}. 
The analysis shares several commonalities with the \ac{DE} of \ac{LDPC} code ensembles designed for joint source and channel coding \cite{fresia2010joint}. 
To proceed with the analysis, and with reference to the \ac{EPC} model, we replace the a priori channel (that introduces a constant number of errors $\omega h$ in $\vecscr$) with a \ac{BSC} with crossover probability $\omega$. The choice is justified by observing that \ac{DE} analysis tracks the evolution of the \ac{BP} message distributions in the limit of large blocklenghts, and by noting that, as $h$ and $n$ grow large, the fraction of errors introduced by the \ac{BSC} concentrates around $\omega$.

We resort to two flavors of \ac{DE}: quantized \ac{DE} \cite{RU01a} and \ac{PEXIT} analysis \cite{ten01,Liva2007:EXIT_GC}.
Quantized \ac{DE}  assumes that the messages exchanged by \acp{VN} and \acp{CN}, as well as the decoder input $L$-values, are uniformly-quantized with sufficiently-fine quantization steps (e.g., we adopted $255$ quantization intervals over a range $[-25,+25]$). The analysis yields accurate \ac{BP} decoding threshold estimates. However, owing to the fine quantization, and the need of tracking a different distribution for each \ac{VN}/\ac{CN} pair in the mother \ac{LDPC} code protograph, the analysis is computationally expensive and can be a bottleneck for a numerical optimization of the mother \ac{LDPC} code protograph. \ac{PEXIT} analysis employs a Gaussian approximation for message distributions, enabling a fast evaluation of the  \ac{BP} decoding threshold. 
In this case, conditioned on the transmitted bit value, all messages and input $L$-values are modeled as Gaussian \acp{r.v.}. Note that the Gaussian approximation is also enforced for the $L$-values associated with the \acp{VN} connected to the a-priori channel. Conditioned on $X=+1$ (allzero codeword assumption), their actual distribution has two mass points, one at $+\Delta$ (with probability $\omega$) and one at $-\Delta$   (with probability $1-\omega$). 
The \ac{PEXIT} analysis follows the steps in \cite{Liva2007:EXIT_GC}, where, in our case, particular attention should be paid to the initialization of the \ac{PEXIT} analysis recursions. 
Denote by $\cbsc(\omega)=1-\Hb(\omega)$ the capacity of a \ac{BSC} with crossover probability $\omega$, and by $\cawgn(E_s/N_0)$ the capacity of a \ac{biAWGN} channel with \ac{SNR} $E_s/N_0$. The \ac{PEXIT} analysis  is initialized by setting the \ac{MI} at the input of type-$i$ \acp{VN}, $i=1,\ldots,h_0$, to $\cbsc(\omega)$, whereas for $i=h_0+1,\ldots, h_0+n_0$ the \ac{MI} at the input of type-$i$ \acp{VN} is initialized to $\cawgn(E_s/N_0)$. The analysis in then carried out via the recursions provided in \cite{Liva2007:EXIT_GC}, and it allows to determine (for fixed $\omega$) the iterative decoding threshold over the \ac{biAWGN} channel, that is the minimum $E_s/N_0$ for which the \ac{MI} values tracked by the \ac{PEXIT} analysis converge to $1$. We denote the threshold value as $\thr(R)$, where we emphasize the dependency on $R$ (and, hence, on $\omega$).

The thresholds predicted by \ac{PEXIT} analysis are close to the ones obtained via quantized \ac{DE} when the rate of the outer code is medium/high. In the low code rate regime, \ac{PEXIT} may provide too optimistic estimates of the \ac{BP} decoding threshold. This effect is illustrated in detail in Section  \ref{sec:PEXITacc}. Owing to this observation, and due to the faster computations entailed by the \ac{PEXIT} analysis with respect to quantized \ac{DE}, \ac{PEXIT} analysis will be used for protograph optimization (provided in Section \ref{sec:design}) in the medium/high (outer code) rate regime, while the use of the slower ---but more accurate--- quantized \ac{DE} analysis will be limited to obtain thresholds at low outer code rates. 

\subsection{Accuracy of the PEXIT Analysis}\label{sec:PEXITacc}
\bgroup
\def\arraystretch{1.15}
\begin{table}[t]
	\caption{\ac{BP} decoding thresholds (\textnormal{d}B) computed by quantized DE and by PEXIT analysis for various rates. Ensemble defined by the base matrix $\bmatonehalfx$ in Example \ref{ex:accuracy}.}
	\centering
	\vspace{-0mm}
	\begin{tabular}{c c c c c c}
		\hline\hline
		{$R$} & $0.5$ & $0.4$ & $0.3$ & $0.2$ & $0.1$ \\ \hline
		$\thr$ dB (quant. \ac{DE}) & $-2.04$  & $-3.40$  & $-4.89$  & $-6.91$  & $-10.27$  \\
		$\thr$ dB ({PEXIT}) & $-2.06$  & $-3.42$  & $-5.05$  & $-7.14$  & $-10.49$  \\ \hline\hline
	\end{tabular}
	\label{tab:DEvsEXIT}
\end{table}
\egroup
\bgroup
\def\arraystretch{1.15}
\begin{table}[t]
	\caption{\ac{BP} decoding thresholds (\textnormal{d}B) computed by quantized DE and by PEXIT analysis for various rates. Ensemble defined by the base matrix $\bmatx{1}{2/3}$ in Example \ref{ex:accuracy2}.}
	\centering
	\vspace{-0mm}
	\begin{tabular}{c c c c c c c}
		\hline\hline
		{$R$} & $0.6$ & $0.5$ & $0.4$ & $0.3$ & $0.2$\\ \hline
		$\thr$ dB (quant. \ac{DE})   &$-0.69$ & $-2.12$  & $-3.43$  & $-4.72$  & $ -6.51$    \\
		$\thr$ dB ({PEXIT})  & $-0.72$& $-2.15$  & $-3.55$  & $-5.00$  & $-7.11$    \\ \hline\hline
	\end{tabular}
	\label{tab:DEvsEXIT23}
\end{table}
\egroup

We consider next two examples that aim at quantifying the accuracy of the \ac{PEXIT} analysis. In particular, we computed the \ac{BP} decoding thresholds of protograph \ac{MN} code ensembles under \ac{PEXIT} analysis, and compare them with the ones derived by quantized \ac{DE}.

\begin{example}\label{ex:accuracy}
	Consider the protograph base matrix \begin{align}
		\bmatonehalfx &= \left(
		\begin{array}{cc|cccc}
			1 &  0 & 1 & 1 & 0 & 0\\
			0 &  1 & 0 & 3 & 0 & 1\\
			2 & 0 & 1 & 1 & 1 & 0\\
			1 & 2 & 1 & 2 & 0 & 0
		\end{array}
		\right).
	\end{align}
	The protograph defines an inner mother \ac{LDPC} code ensemble with rate $\inmR=1/3$ that can be used to construct a protograph \ac{MN} code ensemble with inner code rate $\inR=1/2$. By fixing different values of $\omega$, we obtain different code rates according to \eqref{eq:rates}. Table~\ref{tab:DEvsEXIT} reports the \ac{BP} decoding thresholds for various rates. The values are computed with both quantized \ac{DE} and \ac{PEXIT} analysis. The threshold computed via \ac{PEXIT} analysis is only $0.02$~dB away from the quantized \ac{DE} threshold for $R=0.5$. At lower rates, the gap between the thresholds grows larger: at $R=0.1$, \ac{PEXIT} underestimates the threshold by approx. $0.22$~dB.
\end{example}  

This first example already provides numerical evidence of the accuracy of \ac{PEXIT} when the outer code rate is sufficiently large and of a relative lack of precision at low code rates. A more striking case is provided by the following example.

\begin{example}\label{ex:accuracy2}
	Consider the protograph base matrix 
	\begin{align}
		\bmatx{1}{2/3} &= \left(
		\begin{array}{cc|ccc}
			1 &  0 & 0 & 3 & 1\\
			1 &  1 & 0 & 3 & 0\\
			1 & 2 & 2 & 1 & 0
		\end{array}
		\right).
	\end{align}
	The inner mother \ac{LDPC} code ensemble has rate $\inmR=2/5$. Table~\ref{tab:DEvsEXIT23} shows the \ac{BP} decoding thresholds of the corresponding protograph \ac{MN} code ensemble for various rates. The values are computed with both quantized \ac{DE} and \ac{PEXIT} analysis. As for Example \ref{ex:accuracy}, the \ac{PEXIT} analysis tends to underestimate thresholds. In particular, the accuracy of \ac{PEXIT} analysis strongly deteriorates with decreasing rate. While for a rate of $0.6$ the \ac{PEXIT} threshold estimate is only $0.03$~dB from the quantized \ac{DE} one, for a rate of $0.2$ the gap is around $0.6$~dB. 
\end{example}  

%%%%%%%%%%%%%%%%%%%%%%%%%%%%%%%%%%%%%%%%%%%%%%%%%%%%%%%%%%%%%%%%%%%%%%%%%
%%%%%%%%%%%%%%%%%%%%%%%%%%%%%%%%%%%%%%%%%%%%%%%%%%%%%%%%%%%%%%%%%%%%%%%%%
%%%%%%%%%%%%%%%%%%%%%%%%%%%%%%%%%%%%%%%%%%%%%%%%%%%%%%%%%%%%%%%%%%%%%%%%%

\section{Distance Spectrum Analysis}\label{sec:WEF}

While the \ac{DE} analysis of \ac{LDPC}, \ac{MN}, and in general turbo-like code ensembles provides a useful characterization of the code performance in the so-called waterfall region of the error probability curve, it fails to capture error floor phenomena that may arise at moderate-low error probabilities. Methods that rely on the knowledge of the average weight enumerators of code ensembles are often used to complement \ac{DE} analysis, allowing to discriminate between code ensembles characterized by good minimum distance properties (e.g., ensembles that yield with high probability codes whose minimum distance grows linearly in the block length) and code ensembles with bad minimum distance properties (e.g., ensembles that yield with high probability codes whose minimum distance grows sub-linearly in the block length) \cite{DUR06}. By analyzing the distance properties of a certain code ensemble, it is thus possible to characterize the error floor region of the error probability curve \cite{BM96b}. In this Section, a distance spectrum analysis of protograph \ac{MN} code ensembles is presented. We first derive a union bound on the average block error probability (Section \ref{sec:UB}). The average is here over both the code and the transmitted codeword.  The focus is on \ac{MM} decoding as in Section~\ref{sec:decoding:decoding_metrics}. 
The derivation of the union abound allows identifying the kind of weight enumerator required to analyze the error floor regime. A rigorous derivation of the average weight enumerator is then provided, together with a characterization of the distance properties of code ensembles (Section \ref{sec:AWE}).

\subsection{Union Bound under Mismatched Decoding}\label{sec:UB}

To carry on the derivation of an upper bound on the block error probability under \ac{MM} decoding, we resort to the \ac{EPC} model of Section \ref{sec:MNcodes:equivmodels}. By resorting to the \ac{EPC} setting, the derivation of bounds on the error probability under \eqref{eq:ML3} reduces to the analysis of the error probability under \eqref{eq:ML4}. We first consider transmission with a \ac{MN} code $\code$. The \ac{PEP} is
\begin{equation}
	\PEP(\vecx') = \mathsf{P} \big[ \,p(\vecY|\vecx)Q(\vecScr|\vecw)\leq p(\vecY|\vecx')Q(\vecScr|\vecw')\,\big]. \label{eq:PEP1}
\end{equation}
In \eqref{eq:PEP1}, the codeword transmitted over the communication channel is $\vecx$, and it is the result of the encoding of $\vecw$, where the vector $\vecw$ is transmitted over the a priori channel. The competing codeword is $\vecx'$,  and it is the result of the encoding of $\vecw'$. Note that in \eqref{eq:PEP1} ties are broken in favor of the competing codeword. Owing to the symmetry of the communication and a priori channels, and to the linearity of $\incode$, we assume without loss of generality that $\vecw = (0, 0, \ldots, 0)$ and hence $\vecx=(+1, +1, \ldots, +1)$. Conditioned on $\bm{X}=\vecx$ and $\bm{W}=\vecw$, $\vecScr$ is uniformly distributed over the set of $h$-bit sequences with Hamming weight $\omega h$, whereas $Y_1, Y_2, \ldots, Y_n$ are \ac{i.i.d.} $\sim \mathcal{N}(+1,\sigma^2)$.
We can rewrite \eqref{eq:PEP1} as
\begin{align}
	\PEP(\vecx') 
	&= \prob{\sum_{i=1}^n \ln  \frac{p(Y_i|x_i)}{p(Y_i|x_i')} \leq \sum_{i=1}^h \ln  \frac{Q(\Sscr_i|w_i')}{Q(\Sscr_i|w_i)}}\\
	&= \prob{\textstyle \sum_{i\in \mathcal{D}(\vecx')}  L_i \leq - \sum_{i\in \mathcal{D}(\vecw')} T_i}.
	\label{eq:PEP2}
\end{align}
In \eqref{eq:PEP2}  we made use of
\[
\mathcal{D}(\vecx')=\{i | x_i' \neq x_i\}, \qquad \mathcal{D}(\vecw')=\{i | w_i' \neq w_i\}
\]
whereas 
\[
L_i:= \ln \frac{p(Y_i|0)}{p(Y_i|1)}, \qquad T_i := \ln \frac{Q(\Sscr_i|0)}{Q(\Sscr_i|1)}.
\]
Denote by 
\[
L := \sum_{i\in \mathcal{D}(\vecx')}  L_i\qquad\text{and}\qquad T := \sum_{i\in \mathcal{D}(\vecw')} T_i.
\]
Moreover, let
$a = \hd(\vecw,\vecw')$ and $b = \hd(\vecx,\vecx')$. 
Conditioned on $\bm{X}=\vecx$ and $\bm{W}=\vecw$, we have  
\[
L\sim\mathcal{N}\left(2\frac{b}{\sigma^2},4\frac{b}{\sigma^2}\right).
\]
Recalling $\Delta = \ln [(1-\omega)/\omega]$, we have that $T_i = \Delta$ if $\Sscr_i=0$, whereas $T_i = -\Delta$ if $\Sscr_i=1$, i.e.,
\[
T = (a - E)\Delta - E\Delta = (a -2E)\Delta
\]
where the \ac{r.v.} $E$  follows a hypergeometric distribution with parameters $(h,\omega h, a)$. For the \ac{PEP} we obtain
\begin{equation}
	\PEP(\vecx') = \expect{Q\left(\frac{2b/\sigma^2 + (a -2E)\Delta}{2\sqrt{b}/\sigma}\right)}\label{eq:PEPfin}
\end{equation}
where $Q(x)$ is the well-known Gaussian $Q$-function.
By observing that $\PEP(\vecx')$ depends on $\vecx'$ only through its Hamming distance from $\vecx$, and on the Hamming distance between the corresponding information sequence $\vecw'$ and $\vecw$, we can upper bound the block error probability under \eqref{eq:ML4} as 
\begin{align}
	P_B &\leq 
	\sum_{a = 1}^{h} \sum_{b = 1}^{n} \iowe_{a,b}  \expect{Q\left(\frac{2b/\sigma^2 + (a -2E)\Delta}{2\sqrt{b}/\sigma}\right)} \label{eq:UB}
\end{align}
where $\iowe_{a,b}$ is the input-output weight enumerator of $\incode$. If the code is drawn uniformly at random from a protograph \ac{MN} code ensemble, by averaging over the code ensemble, we obtain the upper bound over the average block error probability
\begin{align}
	\expect{P_B(\code)} &\!\leq \!
	\sum_{a = 1}^{h} \sum_{b = 1}^{n}\! \aiowe_{a,b}  \expect{Q\!\left(\frac{2b/\sigma^2\! +\! (a\! -\!2E)\Delta}{2\sqrt{b}/\sigma}\right)} \label{eq:AUB}
\end{align}
where $\aiowe_{a,b}$ is the average input-output weight enumerator of $\incode$.

\begin{remark} By inspection of \eqref{eq:AUB}, we see that a key role in the block error probability analysis is played by the 
	average input-output weight enumerator of the code ensemble. In particular, numerical evaluation of the \ac{PEP} for various values of $a$ and $b$ reveals that, at low probabilities of error, \eqref{eq:UB} and \eqref{eq:AUB} are dominated by the terms of the (average) input-output weight enumerator associated with small input weight $a$ and small output weight $b$ over a wide range of parameters. More specifically, for very small values of $\omega$ (very low outer code rate $\dmR$) and very low \ac{SNR}, terms of the input-output weight enumerator with small input weight $a$ yield a dominant contribution to the error probability. For a broad region of intermediate values of $\omega$ and of the \ac{SNR}, the terms that contribute mostly to the error probability are associated with small input \emph{and} small output weights. As $\omega$ approaches $1/2$ (the outer code rate $\dmR$ approaches $1$) and the \ac{SNR} grows large, the output weight $b$ gradually takes a predominant role.
\end{remark}

\subsection{Average Input-Output Weight Enumerators}\label{sec:AWE}

Let $\setVN$ be a subset of \acp{VN} in the lifted graph of $\incode$. We assign the value $1$ to each of the \acp{VN} in $\setVN$ and the value $0$ to the \acp{VN} outside $\setVN$. The set $\setVN$ contains $a$  punctured \acp{VN} and $b$ unpunctured ones.
Before deriving the average input-output weight enumerator, we define the \ac{VN} and edge \emph{weight vectors}. 
Define the \ac{VN} weight vector $\weightv=( \weightvj_1, \weightvj_2,\ldots, \weightvj_{h_0+n_0})$, where $\weightvj_j$ is the number of \acp{VN} of type-$j$ in $\setVN$. Recalling that the protograph lifting factor is $\ell$, we have
\begin{align}
	0 &\leq \weightvj_j \leq \liftfac, \text{ for all }  j \in \{1,\ldots,h_0+n_0\} \label{eq:constrtheta0} 
\end{align}
with the constraints 
\begin{align}
	\sum\limits_{j =1}^{h_0} \weightvj_j &=a \label{eq:constrtheta1}\\
	\sum\limits_{j =h_0+1}^{h_0+n_0} \weightvj_j &=b . \label{eq:constrtheta2}
\end{align}
Similarly, define the edge weight vector $\bm{\kp}(\weightv)=(\kp_g)_{g \in \edges}$ where $\kp_g$ is the number of edges of type-$g$ connected to the \acp{VN} in $\setVN$. The \ac{VN} and edge weight vectors are related: for a given $\weightv$, we have  $\kp_g= \weightvj_j$ if $g \in \edgesvj$. Recalling that $n=n_0 \ell$ and $h=h_0 \ell$, the average input-output weight enumerator of the inner \ac{LDPC} code is given by the following lemma.

\begin{lemma}\label{lemma:IOWEF}
	The average number of codewords with input weight $a$ and output weight $b$ in  a code drawn uniformly at random from the protograph \ac{LDPC} code ensemble defined by $\pgraph$ is
	\begin{align}
		\aiowe_{a,b} =&    \sum_{\weightv} \frac{ \prod\limits_{i =1}^{n_0} \coeff\left(   S_i(\bm{z}_i)^\liftfac  ,\bm{z}_i^{\bm{\kp}_i(\weightv)} \right)}{\prod\limits_{j =1}^{h_0+n_0}\binom{\liftfac}{\weightvj_j}^{\dvj-1}}  \label{eq:Eab} \\ 
		\intertext{{where}}
		S_i(\bm{z}_i)=& \frac{1}{2}\left[ \prod\limits_{g \in \edgesci} (1+z_g)+\prod\limits_{g \in \edgesci}  (1-z_g) \right]  \label{eq:Ai}
	\end{align}
	and where  $\bm{\kp}_i(\weightv) = (\kp_g)_{g \in \edgesci}$,   $\kp_g= \weightvj_j$ if $g \in \edgesvj$ and $\bm{z} = (z_g)_{g \in \edges}$, $\bm{z}_i = (z_g)_{g \in \edgesci}$, and $z_g, g \in \edgesci$ are dummy variables. The sum is over the \ac{VN} weight vectors $\weightv=( \weightvj_1, \weightvj_2,\ldots, \weightvj_{h_0+n_0})$ satisfying  \eqref{eq:constrtheta0}-\eqref{eq:constrtheta2}.
\end{lemma}

The proof of Lemma \ref{sec:AWE} is provided in Appendix \ref{appendix:IOWEF}. Lemma \ref{lemma:IOWEF} provides the average  number of codewords with input weight $a$ and output weight $b$  for a
finite block length $n$, and the result can be readily used in \eqref{eq:AUB} to upper bound the average block error probability of a protograph \ac{MN} code ensemble $\left\{\ensP{\pgraph}{\omega}\right\}_{\omega \in (0,1)}$.

To obtain information about the scaling of the minimum distance with the blocklength, it is useful to analyze the normalized logarithmic asymptotic input-output weight distribution for the \ac{LDPC} code ensemble defined by $\pgraph$  for $a=\alpha n$ and $b=\beta n $, which is  defined as
\begin{equation}
	G(\alpha,\beta):= \lim\limits_{n \to \infty}\frac{1}{n}\ln \aiowe_{\alpha n ,\beta n}
\end{equation}
$ 0 \leq \alpha \leq h/n $, $ 0 \leq \beta \leq 1$. The result is provided by the following theorem.

\begin{theorem} \label{theorem:IOG}
	The normalized logarithmic asymptotic input-output weight distribution for the \ac{LDPC} code ensemble defined by $\pgraph$ with input weight $\alpha n$ and output weight $ \beta n$ is 
	\begin{equation} \label{eq:G}
		\begin{split}
		G(\alpha,\beta) = & \frac{1}{n_0} \sum\limits_{i=1}^{n_0} \ln   S_i(\bm{z}^\ast_i)\\  & - \sum\limits_{j=1}^{h_0+n_0}\left[ \frac{\dvj-1}{n_0} H(n_0 \ttheta_j^\star)  +  \ttheta_j^\star \sum\limits_{g \in \edgesvj}  \ln z^\ast_g\right]
		\end{split}
	\end{equation}
	where we recall that $H(p)$ is the natural binary entropy function. 
	The values $z^\ast_g$ for $g \in \edges$,  $\ttheta_j^\star$  for $j \in \{1, \ldots, h_0+n_0\}$ and $\mu_1, \mu_2$ are the solutions of
	\begin{align} 
		z_g \frac{\partial \ln S_i(\bm{z}_i) }{\partial z_g} =&  n_0 \ttheta_j \label{eq:solutions_TS_x}\\
	\intertext{where $g \in \edgesci \cap \edgesvj,  i \in \{1,\ldots,n_0\}, j \in \{1,\ldots,h_0+n_0\}$, and of} 
		(\dvj-1) \ln\left( \frac{\ttheta_j }{\frac{1}{n_0}- \ttheta_j }\right) =& \sum\limits_{g \in \edgesvj} \ln(z_g) + \mu_1 \label{eq:solutions_TS_h}  \\ \intertext{for $j \in \{1,\ldots,h_0\}$,}
		(\dvj-1) \ln\left( \frac{\ttheta_j}{\frac{1}{n_0}- \ttheta_j }\right) =& \sum\limits_{g \in \edgesvj} \ln(z_g) + \mu_2 \label{eq:solutions_TS_n} \\ \intertext{for $j \in \{h_0+1,\ldots,h_0+n_0\}$, and}
		\sum\limits_{j=1}^{h_0} \ttheta_j=&\alpha \label{eq:solutions_TS_mu1} \\
		\sum\limits_{j=h_0+1}^{h_0+n_0} \ttheta_j=&\beta \label{eq:solutions_TS_mu2}
	\end{align}
	where  $S_i(\bm{z}_i)$ is defined in \eqref{eq:Ai}.
\end{theorem}
The proof of Theorem \ref{theorem:IOG} can be found in  Appendix \ref{appendix:IOG}. 
Theorem  \ref{theorem:IOG} shows that the evaluation of $ G(\alpha, \beta) $ 
requires solving $|\edges|+ h_0+n_0+2$ equations with the same number of  variables: $z_g$ ($|\edges|$ variables), $\ttheta_j^\star$ ($h_0+n_0$ variables) and $\mu_1, \mu_2$ (2 variables). 

\medskip

We present the following toy example to clarify the notation.
	\begin{example}
		Consider again the protograph base matrix from Example \ref{ex:proto}, given by 
		\[
		\vB = \left(
		\begin{array}{c|cc}
			1 &  2 & 0\\
			1 &  1 & 2\\
		\end{array}
		\right).
		\]
		We have $h_0=1$ and $n_0=2$. The set of edges in the protograph is $\edges = \{1,2,\ldots, 7\}$.  Thus $|\edges| = 7$. For each 
		\ac{CN} and \ac{VN} in the protograph, we determine the set of edges in the protograph connected to it. We have $\edges_{\cn_1} = \{1,2,3\}, \edges_{\cn_2} = \{4,5,6,7\}, \edges_{\vn_1} = \{1,4\}, \edges_{\vn_2} = \{2,3,5\}, \edges_{\vn_3} = \{6,7\}$. Thus, the 
		\ac{VN} degrees are $\mathsf{d}_{\vn_1}=2, \mathsf{d}_{\vn_2}=3$ and $ \mathsf{d}_{\vn_3}=2  $ . 
		The generating function of the \ac{CN} $\cn_1$ is given by
		\begin{equation}
			S_1(\bm{z}_1)= \frac{1}{2}\left[ \prod\limits_{g \in \{1,2,3\}}
			(1+z_g)+\prod\limits_{g \in \{1,2,3\}}  (1-z_g) \right]
		\end{equation}
		where $\bm{z}_1 =(
		z_1, z_2, z_3)$ and $ z_1, z_2, z_3$ are dummy variables. Similarly, The generating function of the \ac{CN} $\cn_2$ is 
		\begin{equation}
			S_2(\bm{z}_2)= \frac{1}{2}\left[ \prod\limits_{g \in \{4,5,6,7\} } (1+z_g)+\prod\limits_{g \in \{4,5,6,7\}}  (1-z_g) \right]
		\end{equation}
		where  $\bm{z}_2 =(
		z_4, z_5, z_6,  z_7) $ and $ z_4, z_5, z_6, z_7$ are dummy variables.
		For  a fixed integer pair $(a,b)$, we have 
		\begin{align}
			\aiowe_{a,b} =&    \sum_{\weightv} \frac{  \coeff\left(   S_1(\bm{z}_1)^\liftfac  ,\bm{z}_1^{\bm{\kp}_1(\weightv) } \right) \coeff\left(   S_2(\bm{z}_2)^\liftfac  ,\bm{z}_2^{\bm{\kp}_2(\weightv) } \right)}{\binom{\liftfac}{\weightvj_1} \binom{\liftfac}{\weightvj_2}^{2} \binom{\liftfac}{\weightvj_3} }
		\end{align}
		where
		the sum is over the \ac{VN} weight vectors $\weightv = ( \weightvj_1, \weightvj_2, \weightvj_3) $ satisfying 
		$0 \leq \weightvj_1, \weightvj_2, \weightvj_3 \leq \liftfac, 
		\weightvj_1 = a$ and $ \weightvj_2 + \weightvj_3 = b$.
		For each  \ac{VN} weight vectors $\weightv = ( \weightvj_1, \weightvj_2, \weightvj_3) $, we can compute the edge weight vector $\bm{\kp}(\weightv) =( \kp_1, \kp_2, \ldots,\kp_7)  $. Since  $\kp_g= \weightvj_j$ if $g \in \edgesvj$, we obtain for this example, $\bm{\kp}(\weightv) =(\weightvj_1,  \weightvj_2, \weightvj_2, \weightvj_1, \weightvj_2, \weightvj_3, \weightvj_3)$. Further, for each \ac{CN} $\cn_i$ in the protograph, we determine 
		$\bm{\kp}_i(\weightv) $ which contains the weights of the edges connected to it. Formally,  $\bm{\kp}_i(\weightv) = (\kp_g)_{g \in \edgesci}$. In this case, we have $\bm{\kp}_1(\weightv) =(\kp_1, \kp_2, \kp_3 ) = ( \weightvj_1,  \weightvj_2, \weightvj_2) $ and $\bm{\kp}_2(\weightv) =(\kp_4, \kp_5, \kp_6. \kp_7 ) = ( \weightvj_1, \weightvj_2, \weightvj_3, \weightvj_3)$. The evaluation of the normalized logarithmic asymptotic input-output weight distribution  $ G(\alpha,\beta) $ requires solving a system of equation with $|\edges| + h_0+ n_0 +2 =12$  equations with the same number of unknowns ($ z_1, z_2, z_3, z_4, z_5, z_6, z_7, \ttheta_1, \ttheta_2, \ttheta_3, \mu_1, \mu_2$).
\end{example}

\medskip

The following lemma follows the approach of \cite{paolini_growthrate_2016}, and it can reduce the dimension of the system of equations, hence simplifying the calculation.
\begin{lemma}\label{lemma:reduce}
	Let $u, v $  be two edges in $\edges$. If $u$ and $v$ are connected to the same \ac{VN}-\ac{CN} pair in the protograph, then $z^\ast_u=z^\ast_v$. 
	\begin{proof}
		The function $S_i(\bm{z}_i)$ in \eqref{eq:Ai} is symmetric in the variables $z_g$, $g \in \edgesci$. 
		Thus, for the system of equations in Theorem \ref{theorem:IOG}, if there is a solution with $z^\ast_u=\theta_1, z^\ast_v=\theta_2$ then another solution exists with $z^\ast_u=\theta_2, z^\ast_v=\theta_1$ (all the other variables being unchanged). Since the solutions $z^\ast_g, g \in \edges$ are unique, we have $\theta_1=\theta_2$.
	\end{proof}
\end{lemma}

\begin{remark}\label{rem:badens}
	The derivation of the normalized logarithmic asymptotic input-output weight distribution for the \ac{LDPC} code ensemble defined by $\pgraph$ allows to distinguish between two behaviors. Suppose that $G(\alpha,\beta)$ is strictly positive for  $0<\alpha<\xi$ and $0<\beta<\xi$, where $\xi$ is an arbitrarily small positive constant. In this case, an exponentially large number of codewords with input weight $a=\alpha n$ and output weight $b=\beta n$ is expected, with $\alpha$ and $\beta$ small compared to $n$. According to \eqref{eq:AUB}, ensembles displaying this behavior will be characterized by poor error floor performance since $\aiowe_{a,b}$ will be large for small $a,b$. We will refer to ensembles possessing this property as \emph{bad ensembles}. On the contrary, we refer to ensembles for which there exists a positive $\xi$ such that $G(\alpha,\beta)$ is strictly negative for $0<\alpha<\xi$ and $0<\beta<\xi$ as \emph{good ensembles}. 
\end{remark}

We provide next two examples of ``good'' and ``bad'' code ensembles, according to the definition of Remark \ref{rem:badens}.

\begin{figure*}
	\begin{center}
		\subfloat[]{\includegraphics[width = 7.5cm]{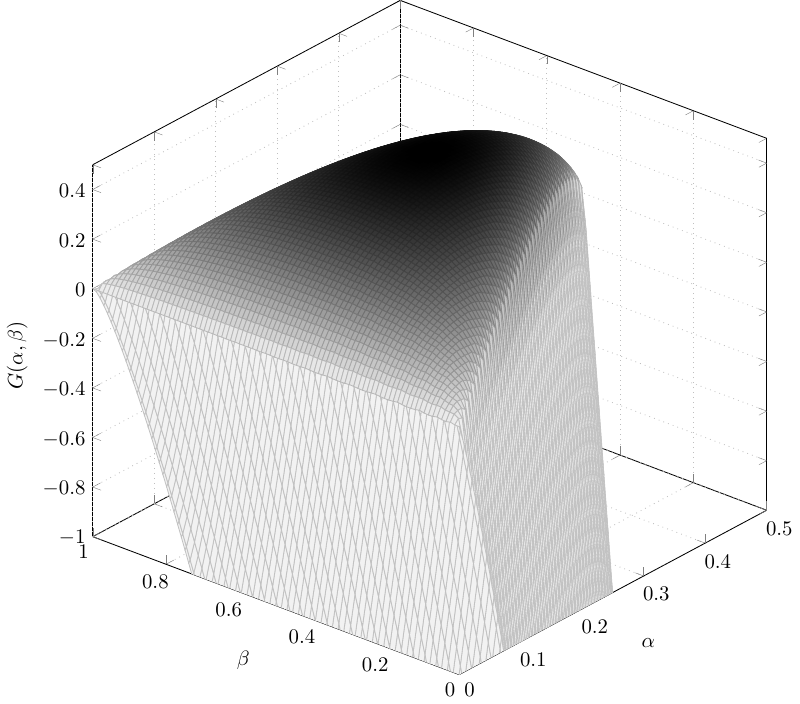}}\hspace{0.5cm}
		\subfloat[]{\includegraphics[width = 7.5cm]{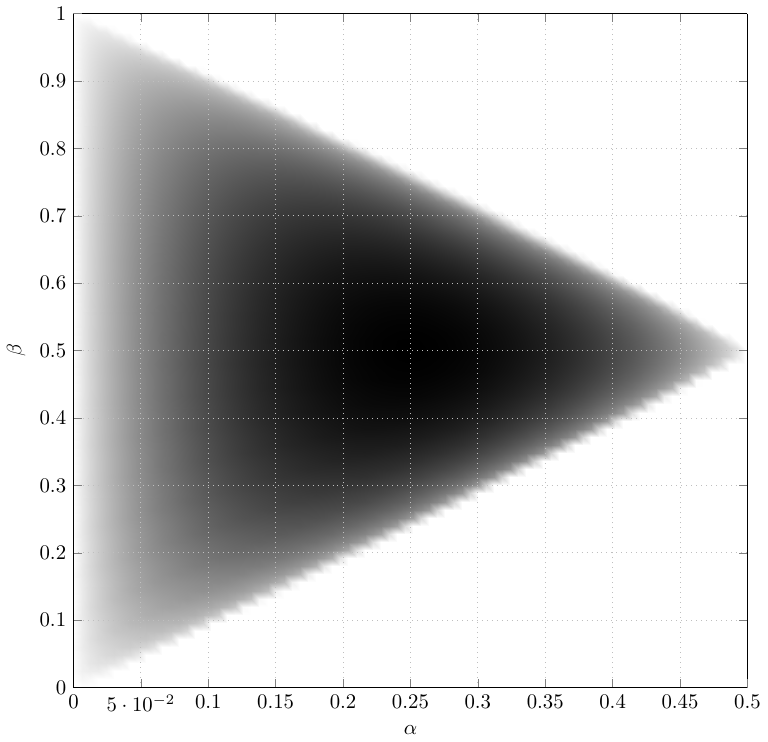}}
	\end{center}
	\caption{Normalized logarithmic asymptotic input-output weight distribution (a) for the \ac{LDPC} code ensemble of Example \ref{ex:bad}. The view from the top is given in (b), where the white color denotes points where $G(\alpha,\beta)$ is negative.}
	\label{fig:ex:bad}
\end{figure*}

\begin{example}\label{ex:bad}
	Consider the protograph base matrix \begin{align}
		\vB &= \left(
		\begin{array}{c|cc}
			1 &  1 & 1\\
			1 &  1 & 1\\
		\end{array}
		\right).
	\end{align}
	The protograph defines an inner mother \ac{LDPC} code ensemble with rate $\inmR=1/3$ that can be used to construct a protograph \ac{MN} code ensemble with inner code rate $\inR=1/2$. Figures \ref{fig:ex:bad}(a) and \ref{fig:ex:bad}(b) depict the normalized logarithmic asymptotic input-output weight distribution. We observe that $G(\alpha,\beta)$ is positive for $\alpha>0$ and $\beta>0$, $\alpha$ and $\beta$ small. Hence, according to Remark \ref{rem:badens}, the ensemble is  ``bad''.
\end{example}

\begin{figure*}
	\begin{center}
		\subfloat[]{\includegraphics[width = 7.5cm]{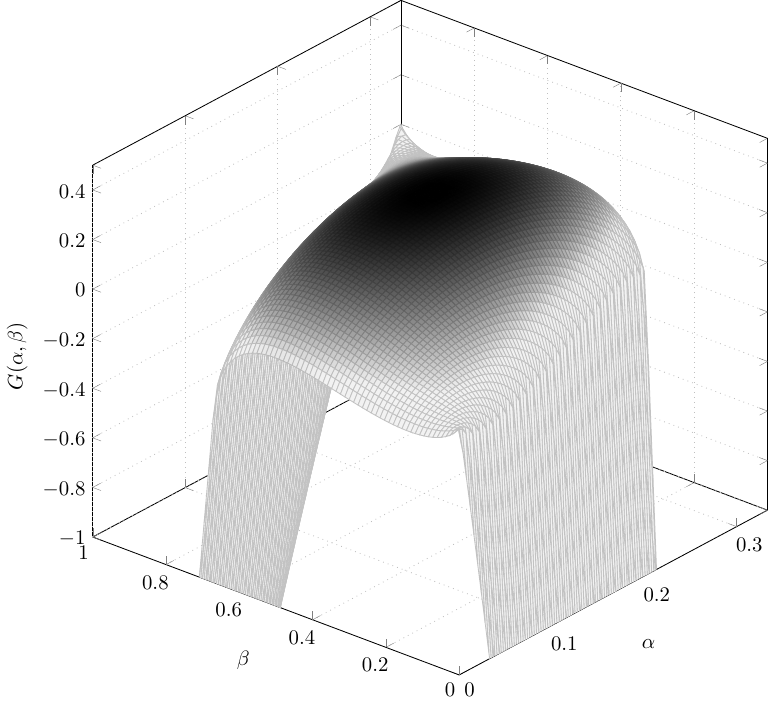}}\hspace{0.5cm}
		\subfloat[]{\includegraphics[width = 7.5cm]{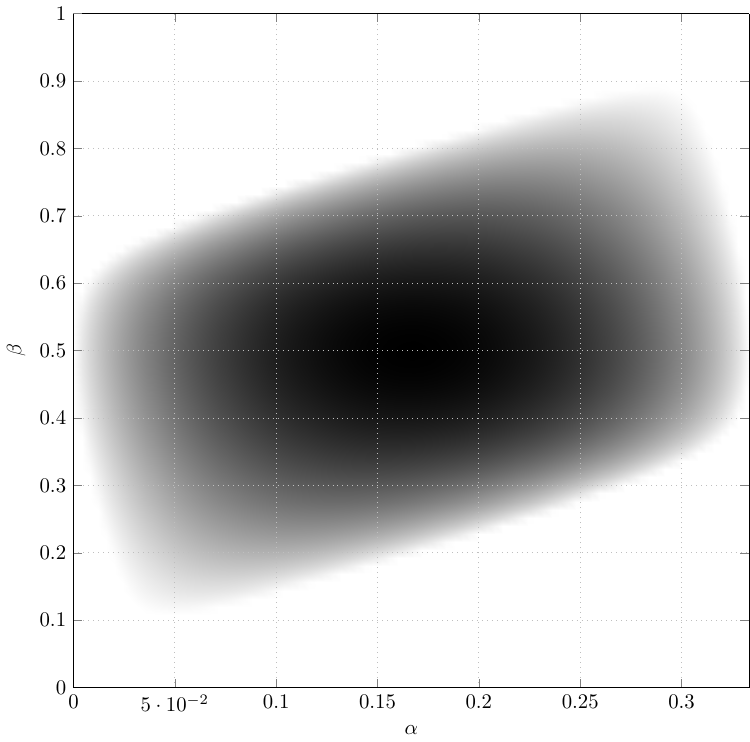}}
	\end{center}
	\caption{Normalized logarithmic asymptotic input-output weight distribution (a) for the \ac{LDPC} code ensemble of Example \ref{ex:good}. The view from the top is given in (b), where the white color denotes points where $G(\alpha,\beta)$ is negative.}
	\label{fig:ex:good}
\end{figure*}

\begin{example}\label{ex:good}
	Consider the protograph base matrix \begin{align}
		\vB &= \left(
		\begin{array}{c|ccc}
			1 &  1 & 1 & 1\\
			1 &  1 & 1 & 1\\
			1 &  1 & 1 & 1\\
		\end{array}
		\right).
	\end{align}
	The protograph defines an inner mother \ac{LDPC} code ensemble with rate $\inmR=1/4$ that can be used to construct a protograph \ac{MN} code ensemble with inner code rate $\inR=1/3$. Figures \ref{fig:ex:good}(a) and \ref{fig:ex:good}(b) depict the normalized logarithmic asymptotic input-output weight distribution. We observe that $G(\alpha,\beta)$ is negative for $\alpha>0$ and $\beta>0$ and $\alpha$ and $\beta$ small. Hence, according to Remark \ref{rem:badens}, the ensemble is  ``good''.
\end{example}

%%%%%%%%%%%%%%%%%%%%%%%%%%%%%%%%%%%%%%%%%%%%%%%%%%%%%%%%%%%%%%%%%%%%%%%%%
%%%%%%%%%%%%%%%%%%%%%%%%%%%%%%%%%%%%%%%%%%%%%%%%%%%%%%%%%%%%%%%%%%%%%%%%%
%%%%%%%%%%%%%%%%%%%%%%%%%%%%%%%%%%%%%%%%%%%%%%%%%%%%%%%%%%%%%%%%%%%%%%%%%

\section{Code Design}\label{sec:design}

A first step in the design of \ac{MN} codes is the identification of a suitable inner \ac{LDPC} code protograph ensemble. For this purpose,  the iterative threshold $\thr$ can be used as cost function to be minimized. In particular, suppose we are interested in finding an inner mother code protograph that allows to operate close to capacity over a range $\left[\rmin,\rmax\right]$ of rates $R$, i.e., over a range $\left[\omin,\omax\right]$ of values of $\omega$.

The protograph search begins by fixing the protograph parameters $h_0, n_0$, which define the inner code rate $\inR$.  A set of target rates $\rset \subset \left[\rmin,\rmax\right]$   is then selected. For each target rate $R\in\rset$ we can derive the rate of the \ac{DM} as $\dmR = R/\inR$, out of which the \ac{DM} parameter $\omega$ is obtained. We make use of the following definition.

\begin{definition}
	Consider a protograph $\pgraph$. We define the \ac{WCL} 
	\begin{equation}
			\WCL(\pgraph,\rset) := \max_{R \in \rset} \left[\thr(R) - \cawgn^{-1}(R)\right].\label{eq:WCL}
	\end{equation}
	The \ac{WCL}  is the maximum gap between the protograph iterative decoding threshold and the \ac{biAWGN} channel Shannon limit for the rates in $\rset$.
\end{definition}

The \ac{WCL} provides a measure of the capability of \ac{MN} codes constructed from the inner \ac{LDPC} protograph ensemble specified by $\pgraph$ to approach the Shannon limit for different choices of the \ac{DM} parameter $\omega$ (and, hence, over the corresponding range of code rates).
A search for the protograph with parameters $h_0, n_0$ that minimizes the \ac{WCL} in \eqref{eq:WCL} can be carried out, for example, via differential evolution \cite{shokrollahi2005design}. We provide next some examples of application to the design of protograph-based \ac{MN} code ensemble families addressing different rate regimes. In all examples, the search space was limited by setting the maximum number of parallel edges between protograph \ac{VN}/\ac{CN} pairs (i.e., the value of the base matrix elements) to $3$.

\begin{figure}
	\centering
	\includegraphics[width=\columnwidth]{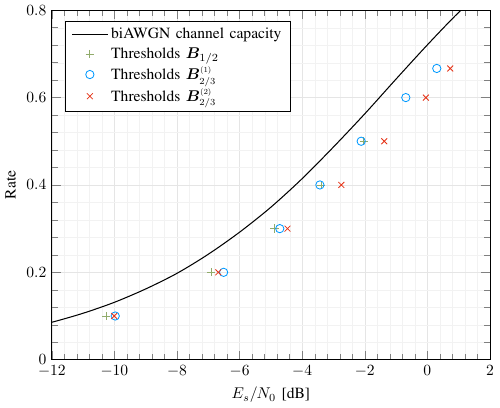}
	\caption{Iterative decoding thresholds computed for the \ac{MN} code ensembles defined by the base matrices described in Example \ref{ex:12}, Example \ref{ex:23}, and in Example \ref{ex:23b}.}
	\label{fig:thresholds}
\end{figure}

\begin{example}\label{ex:12}
	Consider a code rate range $[0.1, 0.5]$. An \ac{MN} code family addressing this range can be derived from an inner $\inR=1/2$ code, i.e., the mother code has rate $\inmR=1/3$. We search for protographs with $6$ \acp{VN} and $4$ \acp{CN}, minimizing the \ac{WCL} over $\rset=\{0.1, 0.3, 0.5\}$. We obtain the base matrix
	\begin{align}
		\bmatonehalfx &= \left(
		\begin{array}{cc|cccc}
			1 &  0 & 1 & 1 & 0 & 0\\
			0 &  1 & 0 & 3 & 0 & 1\\
			2 & 0 & 1 & 1 & 1 & 0\\
			1 & 2 & 1 & 2 & 0 & 0
		\end{array}
		\right)
	\end{align}
	where the first two columns are associated with punctured \acp{VN}.
	Note that the base matrix is the same as in Example \ref{ex:accuracy}. The iterative decoding thresholds (obtained for a small sample of the possible rates) are depicted in Figure \ref{fig:thresholds}.
\end{example}

\begin{example}\label{ex:23}
	Consider a code rate range $[0.1, 0.666]$. We fix the inner code rate to $\inR=2/3$ and the dimensions of the base matrix to $3 \times 5$, where the first two columns are associated with punctured \acp{VN}. We search for protographs minimizing the \ac{WCL} over $\rset=\{0.1, 0.3, 0.666\}$. We obtain the base matrix
	\begin{align}
		\bmatx{1}{2/3} &= \left(
		\begin{array}{cc|ccc}
			1 &  0 & 0 & 3 & 1\\
			1 &  1 & 0 & 3 & 0\\
			1 & 2 & 2 & 1 & 0
		\end{array}
		\right).
	\end{align}
	The iterative decoding thresholds (again, obtained for a small sample of the possible rates) are depicted in Figure \ref{fig:thresholds}.
\end{example}

Remarkably, the iterative decoding thresholds for both ensembles of Examples \ref{ex:12} and \ref{ex:23} (displayed in Figure \ref{fig:thresholds}) are within $1$ dB from the Shannon limit over a wide range of rates. 

For both designs in Examples \ref{ex:12} and \ref{ex:23}, we did not restrict the search to protographs yielding good ensembles (see Remark~\ref{rem:badens}). By studying the input-output weight enumerators of the inner protograph \ac{LDPC} code ensembles defined by the base matrices Examples \ref{ex:12} and \ref{ex:23}, we can observe that both base matrices result in bad ensembles. Hence, we should expect codes constructed from these base matrices to exhibit (relatively) high error floors. Evidence in the form of numerical results will be provided in Section \ref{sec:results}. 

We modify the protograph search by including a check to distinguish between protographs yielding good and bad ensembles, according to the criterion introduced in Remark \ref{rem:badens}. By doing so, bad ensembles can be discarded during the protograph search. An example of code ensemble obtained using this approach is provided next.

\begin{example}\label{ex:23b}
	Consider a code rate range $[0.1, 0.666]$, an inner $\inR=2/3$ code, and a base matrix of dimensions $3 \times 5$. We search for protographs minimizing the \ac{WCL} over $\rset=\{0.1, 0.3, 0.666\}$ excluding bad ensembles. We obtain the base matrix
	\begin{align}
		\bmatx{2}{2/3} &= \left(
		\begin{array}{cc|ccc}
			3 &  3 & 3 & 0 & 0 \\
			0 &  1 & 3 & 1 & 0\\
			1 & 0 & 2 & 0 & 1
		\end{array}
		\right)
	\end{align}
	where the first two columns are associated with punctured \acp{VN}. The base matrix above describes a rate-$2/5$ inner mother \ac{LDPC} code, which can be obtained by concatenating an outer rate-$2/3$ \ac{LDPC} code with regular \ac{VN} degree $3$ and \ac{CN} degree $9$, with an inner low-density generator-matrix code. The construction is reminiscent of the \ac{LDPC} codes adopted by the 3GPP New Radio (5G) standard \cite{Richardson5G} and of the Raptor-based \ac{LDPC} code class introduced in \cite{Divsalar14:ProtoRaptor}. The iterative decoding thresholds for a small set of the possible rates are depicted in Figure \ref{fig:thresholds}. Observe that the thresholds from the constrained optimization result in a slight loss for high rates compared to the unconstrained case of Example \ref{ex:23}.
\end{example}

%%%%%%%%%%%%%%%%%%%%%%%%%%%%%%%%%%%%%%%%%%%%%%%%%%%%%%%%%%%%%%%%%%%%%%%%%
%%%%%%%%%%%%%%%%%%%%%%%%%%%%%%%%%%%%%%%%%%%%%%%%%%%%%%%%%%%%%%%%%%%%%%%%%
%%%%%%%%%%%%%%%%%%%%%%%%%%%%%%%%%%%%%%%%%%%%%%%%%%%%%%%%%%%%%%%%%%%%%%%%%

\section{Numerical Results}\label{sec:results}

Numerical results on the performance of protograph \ac{MN} codes over the \ac{biAWGN} channel are provided next. The results are obtained via Monte Carlo simulations with a maximum of $100$ \ac{BP} decoding iterations. All the codes used in the simulations are obtained by lifting the inner \ac{LDPC} code protograph by a circulant version of the \ac{PEG} algorithm \cite{HEA05}.

\begin{figure}[t]
	\centering
	\includegraphics[width=\columnwidth]{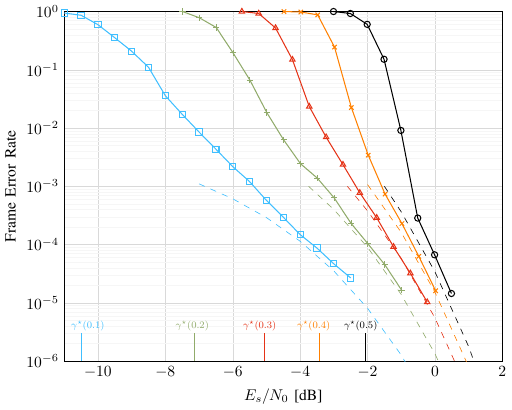}
	\caption{\ac{FER} vs. $E_s/N_0$ (in dB) for length-$1200$ protograph \ac{MN} codes with inner mother code base matrix $\bmatonehalfx$ and for rates $R=0.5$ (black), $R=0.4$ (orange), $R=0.3$ (red), $R=0.2$ (green), and $R=0.1$ (blue). The TUB on the block error probability for each rate is provided (dashed lines) as well as the corresponding iterative decoding thresholds.}
	\label{fig:FER_UB}
\end{figure}

The first set of simulation results confirms the tightness of the union bound in \eqref{eq:UB} at low error probabilities.
For this purpose, we designed a length-$1200$ \ac{MN} codes based on the protograph of Example \ref{ex:12}. The maximum code rate $R=1/2$ is obtained for $\omega=1/2$. The performance for various code rates is depicted in Figure \ref{fig:FER_UB}, in terms of \ac{FER} vs. \ac{SNR}, together with the iterative decoding thresholds of the corresponding \ac{MN} code ensembles  $\ensP{\pgraph}{\omega}$. On the same chart, we depict the union bound \eqref{eq:UB}, truncated to the contributions given by codewords with small input/output weights. Owing to the moderate block length of the code, we resort to an enumeration of the lower tail of the inner code distance spectrum by means of the efficient algorithm introduced in \cite{Hu2004a}. The \ac{TUB} provides an excellent prediction of the \ac{FER} at low error probabilities, confirming the validity of the approach in Section \ref{sec:AWE} for the analysis of the error floor performance.\footnote{Note that the \ac{TUB} simply provides an estimate of the frame error rate at sufficiently low error probabilities. The wording ``error floor'', which is widely used in this context, can be misleading---the frame error rate curves do not ``flatten'' on an actual floor, but rather follow a gentle slope.} Interestingly, the \acp{TUB} indicate at large \ac{SNR} a diminishing return in coding gain when the rate of the outer \ac{CC} is reduced, whereas the coding gains at moderate \ac{FER} are more sizable. In accordance with the analysis of the normalized logarithmic asymptotic input-output weight distribution of the inner \ac{LDPC} code ensemble, the codes show a poor performance at moderate-low error rates. Recall that the base matrix of Example~\ref{ex:12} defines a bad ensemble.
We can see that, at the highest code rate ($R=1/2$) the \ac{TUB} is met at a $\mathsf{FER}\approx 3 \times 10^{-4}$, signaled by a visible change in the slope of the curve. The change in slope is less abrupt at the lower code rates, where the effect of low-input / low-output weight error patterns is causing a performance degradation already at relatively high error rates. At rate $1/10$, it is almost impossible to distinguish the waterfall region.

\begin{figure}[t]
	\centering
	\includegraphics[width=\columnwidth]{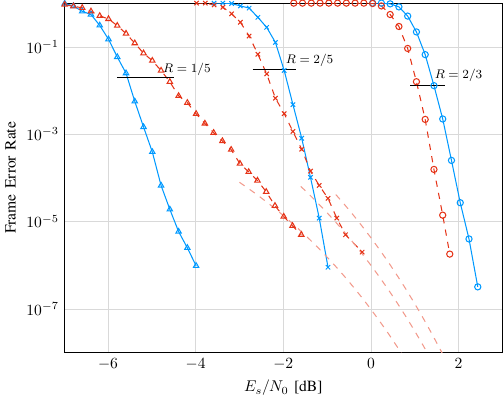}
	\caption{\ac{FER} vs. $E_s/N_0$ dB for length-$1800$ protograph \ac{MN} codes.  with inner mother code base matrix $\bmatxcap{1}{2/3}$ (Example \ref{ex:23}, in red, dashed lines) and $\bmatxcap{2}{2/3}$ (Example \ref{ex:23b}, in blue, solid lines). The performance is provided for the \ac{MN} code family of Example \ref{ex:23}, with base matrix $\bmatxcap{1}{2/3}$, for rates $R=1/5$, $R=2/5$, and $R=2/3$. Similarly, the performance is provided for the \ac{MN} code family of Example \ref{ex:23b}, with base matrix $\bmatxcap{2}{2/3}$, for rates $R=1/5$, $R=2/5$, and $R=2/3$. TUBs for the MN code family with base matrix $\bmatxcap{1}{2/3}$ are provided (light-red, dashed lines) for rates $R=1/5$, $R=2/5$, and $R=2/3$ (left to right).}
	\label{fig:FER_short}
\end{figure}

The performance of \ac{MN} codes of different rates and with block length $1800$ are provided in Figure \ref{fig:FER_short}, in terms of \ac{FER} vs. \ac{SNR}. The chart includes results of for two families of \ac{MN} codes constructed from the \ac{MN} code ensembles of Examples~\ref{ex:23}~and~\ref{ex:23b}. Note that while a fine granularity of code rates can be achieved by suitably choosing the \ac{DM} parameter $\omega$, for the simulations, only three rates were sampled, i.e., the maximum code rate allowed by the ensembles ($R=2/3$), a low rate ($R=1/5$) and an intermediate rate ($R=2/5$). The \ac{MN} codes defined by the ensemble of Example \ref{ex:23b} show superior performance compared to the codes defined by the ensemble of Example \ref{ex:23}, at low error rates. In particular, for the code derived from $\bmatx{1}{2/3}$  the algorithm of \cite{Hu2004a} was able to find low-weight codewords (e.g., with input weight $4$ and output weight $17$) that are responsible of the poor performance of the $R=1/5$ and $R=2/5$ codes at moderate-low error rates. As observed for the codes defined by the protograph of Example \ref{ex:12}, the flooring phenomenon is particularly severe at the lower code rates, where the error rate curve does not show a proper waterfall region.  At the highest rate ($R=2/3$), we may expect the actual \ac{FER} to approach the respective \ac{TUB} at error rates below $10^{-8}$. We applied the search algorithm of \cite{Hu2004a} to the code defined by the base matrix $\bmatx{2}{2/3}$. The search returned only codewords of relatively large weight (e.g., input weight $27$, output weight $95$), resulting in \ac{TUB} \ac{FER} predictions that are well below the \ac{FER} values depicted in Figure \ref{fig:FER_short}. Note finally that, in the simulated error rate regime, the rate-$2/3$ the code defined by the base matrix $\bmatx{1}{2/3}$ outperforms its counterpart based on $\bmatx{2}{2/3}$. This fact is consistent with the analysis of Figure \ref{fig:thresholds}: the performance in the waterfall region is still dominated by the asymptotic iterative decoding threshold. While the simulation results do not allow us to explore the error floor performance for the rate-$2/3$ codes, owing to the \ac{TUB} analysis, we may still expect the code derived from $\bmatx{2}{2/3}$ to outperform the code derived from $\bmatx{1}{2/3}$, at very low error rates.
	These results confirm that the criterion identified in Remark \ref{rem:badens} to differentiate good and bad ensembles (from an error floor perspective) can be used, at an early design stage, to discard protographs that yield codes with poor performance at low error rates.

\begin{figure}[t]
	\centering
	\includegraphics[width=\columnwidth]{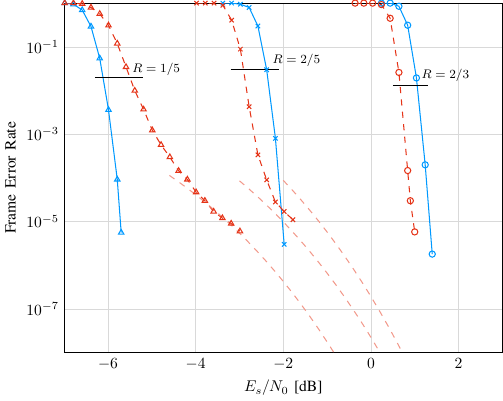}
	\caption{\ac{FER} vs. $E_s/N_0$ dB for length-$9000$ protograph \ac{MN} codes.  with inner mother code base matrix $\bmatxcap{1}{2/3}$ (Example \ref{ex:23}, in red, dashed lines) and $\bmatxcap{2}{2/3}$ (Example \ref{ex:23b}, in blue, solid lines). The performance is provided for the \ac{MN} code family of Example \ref{ex:23}, with base matrix $\bmatxcap{1}{2/3}$, for rates $R=1/5$, $R=2/5$, and $R=2/3$. Similarly, The performance is provided for the \ac{MN} code family of Example \ref{ex:23b}, with base matrix $\bmatxcap{2}{2/3}$, for rates $R=1/5$, $R=2/5$, and $R=2/3$. TUBs for the MN code family with base matrix $\bmatxcap{1}{2/3}$ are provided (light-red, dashed lines) for rates $R=1/5$, $R=2/5$, and $R=2/3$ (left to right).}
	\label{fig:FER_long}
\end{figure}

The observations derived from Figure \ref{fig:FER_short} are confirmed at larger block lengths. In particular, Figure \ref{fig:FER_long} reports the performance of \ac{MN} codes with block length $9000$, for the same ensembles of Example \ref{ex:23} and of Example \ref{ex:23b}. Due to the relatively large blocklength, all curves are relatively steep, at moderate-high error rates. For the code based on $\bmatx{1}{2/3}$, a drastic change in slope takes place at $\mathsf{FER}\approx 10^{-5}$, for both $R=1/5$ and $R=2/5$. Here, the simulation results approach the respective \ac{TUB} predictions. At the lower rate ($R=1/5$), the waterfall performance of the code from the ensemble of Example \ref{ex:23} is again hindered by the emergence of a high error floor. As for the blocklength-$1800$ case, at the highest rate ($R=2/3$) we may expect the actual \ac{FER} to approach the respective \ac{TUB} at error rates below $10^{-8}$. Again, for the code based on $\bmatx{2}{2/3}$ the \ac{TUB} analysis returned \ac{FER} predictions well below the values depicted in the Figure \ref{fig:FER_long}. As for the blocklength-$1800$ case, in the simulated error rate regime, the rate-$2/3$ code defined by the base matrix $\bmatx{1}{2/3}$ outperforms its counterpart based on $\bmatx{2}{2/3}$ --- recall that the performance in the waterfall region is dominated by the iterative decoding threshold. As before, while the simulation results do not allow us to explore the error floor performance for the rate-$2/3$ codes, owing to the \ac{TUB} analysis, we may still expect the code derived from $\bmatx{2}{2/3}$ to outperform the code derived from $\bmatx{1}{2/3}$, at very low error rates.

%%%%%%%%%%%%%%%%%%%%%%%%%%%%%%%%%%%%%%%%%%%%%%%%%%%%%%%%%%%%%%%%%%%%%%%%%
%%%%%%%%%%%%%%%%%%%%%%%%%%%%%%%%%%%%%%%%%%%%%%%%%%%%%%%%%%%%%%%%%%%%%%%%%
%%%%%%%%%%%%%%%%%%%%%%%%%%%%%%%%%%%%%%%%%%%%%%%%%%%%%%%%%%%%%%%%%%%%%%%%%

\section{Concluding Remarks}\label{sec:conclusions}

Protograph MacKay-Neal (MN) codes have been introduced and analyzed. The code construction relies on the concatenation of  an inner, protograph-based low-density parity-check (LDPC) code with an outer 
constant composition (CC) code. The outer code encoder acts as a distribution matcher (DM) that enables changing the rate of the MN code by simply changing the Hamming weight of the CC codewords. Noting that the resulting concatenation defined a nonlinear block code, an equivalent communication model is introduced. The equivalent model allows analyzing the performance of MN codes by studying the error probability of the inner (linear) LDPC code, with transmission that takes place in parallel over the communication channel, and over a suitably defined binary symmetric channel. A density evolution analysis is provided, and it is complemented by a characterization of the distance properties of the code ensemble. The distance spectrum analysis serves to discriminate between ensembles that will originate codes with high error floors, and ensembles yielding codes with low error floors. A code design technique is proposed. The accuracy of the analysis and the validity of the code design method are confirmed by Monte Carlo simulations. Examples of code designs are provided, showing how the use of a single LDPC code ensemble allows operating within $1$ dB from the Shannon limit over a wide range of code rates, where the code rate is selected by tuning the DM parameters. By enabling rate flexibility with a constant blocklength, and with a fixed LDPC code as inner code, the construction provides an appealing solution for very high-throughput wireless (optical) links that employ binary-input modulations. 

\section*{Acknowledgment}
The authors would like to thank Prof. Gerhard Kramer (Technical University of Munich) for his constructive suggestions and the anonymous reviewers for their insightful comments that helped improve the final version of this paper.

%%%%%%%%%%%%%%%%%%%%%%%%%%%%%%%%%%%%%%%%%%%%%%%%%%%%%%%%%%%%%%%%%%%%%%%%%
%%%%%%%%%%%%%%%%%%%%%%%%%%%%%%%%%%%%%%%%%%%%%%%%%%%%%%%%%%%%%%%%%%%%%%%%%
%%%%%%%%%%%%%%%%%%%%%%%%%%%%%%%%%%%%%%%%%%%%%%%%%%%%%%%%%%%%%%%%%%%%%%%%%

\appendices

\section{Proof of Lemma \ref{lemma:IOWEF}} \label{appendix:IOWEF}
Consider the Tanner graph of an \ac{LDPC} code drawn uniformly at random from the ensemble defined by a protograph $\pgraph$. We randomly choose a set $\setVN$ of $a$ punctured \acp{VN}  and $b$ unpunctured ones.  We assign the value one to each \ac{VN} in $\setVN$ and zero to the \acp{VN} outside $\setVN$. The edges connected to a \ac{VN} $\vn$ are assigned the value  chosen for $\vn$. For a given $\weightv$, each $\vn_j \in \cV$ has $\weightvj_j$ replicas in $\setVN$. Since there are $\liftfac$ copies of each \ac{VN} type in the lifted graph, the number of $\ac{VN}$ sets with weight vector $\weightv$ is given by 
\begin{equation}
	N_{\vn}(\weightv)=\prod\limits_{j =1}^{h_0+n_0} \binom{\liftfac}{\weightvj_j}.
\end{equation}
Since $\kp_g= \weightvj_j$ if $g \in \edgesvj$, the number of edge sets with weight vector $\bm{\kp}(\weightv)$ is 
\begin{equation}
	\begin{aligned}
		N_{e}(\bm{\kp}(\weightv))=&  \prod\limits_{g \in \edges}\binom{\liftfac}{\kp_g}\!=\!\prod\limits_{j=1}^{h_0+n_0} \prod\limits_{g \in \edgesvj}\binom{\liftfac}{\weightvj_j}\!=\!\prod\limits_{j=1}^{h_0+n_0}\binom{\liftfac}{\weightvj_j}^{\dvj}.
	\end{aligned}
\end{equation}
Let $ N_{\cn}(\bm{\kp}(\weightv))$ be the number of configurations with edge   weight vector $\bm{\kp}(\weightv)$ such that all \acp{CN} are satisfied. 
A  \ac{CN} is satisfied if the sum of the bits assigned to its connected edges is zero.
Consider a \ac{CN} of type $i$. The number of configurations for which the \ac{CN} is satisfied is tracked by the generating function
\begin{equation}
	S_i(  \bm{z}_i)=  \!\! \sum\limits_{\substack{\bm{c} \in  \{0,1\}^{\dci}\\ \wH(\bm{c}) \text{ is even}  }} \!\!\! \bm{z}_i^{\bm{c}} =
	\frac{1}{2}\left[ \prod\limits_{g \in \edgesci} \left(1+z_g \right)+ \prod\limits_{g \in \edgesci}  \left(1-z_g\right) \right].
\end{equation}
Considering all \ac{CN} types, and that there are $ \liftfac$ \acp{CN} of each type, we obtain
\begin{equation}
	N_{\cn}(\bm{\kp}(\weightv))=  \prod\limits_{i =1}^{n_0} 
	\coeff\left(  S_i(\bm{z}_i)^\liftfac  ,\bm{z}_i^{\bm{\kp}_i(\weightv)} \right).
\end{equation}
The proof is completed by observing that
\begin{equation}
	\aiowe_{a,b} =    \sum_{\weightv}   \frac{N_{\vn}(\weightv) N_{\cn}(\bm{\kp}(\weightv))}{ N_{e}(\bm{\kp}(\weightv))}
\end{equation}
where the sum is over the \ac{VN} weight vectors $\weightv=( \weightvj_1, \weightvj_2,\ldots, \weightvj_{h_0+n_0})$ satisfying  \eqref{eq:constrtheta0}-\eqref{eq:constrtheta2}.\qed

\section{Proof of Theorem  \ref{theorem:IOG}} \label{appendix:IOG}
From Lemma \ref{lemma:lemmaHayman}, and recalling that $\liftfac= n/n_0$, we have
\begin{equation}\label{eq:coeffAi}
	\begin{split}
	\coeff\left(  S_i(\bm{z}_i)^{\frac{n}{n_0}}  ,\bm{z}_i^{n\bm{\tk}_i(\bm{\ttheta})} \right) 
	\dot=& \exp\Bigg\{n \Bigg[  \frac{1}{n_0}\ln S_i(\bm{z}^\ast_i)\\  &- \sum\limits_{g \in \edgesci}\tk_g \ln z^\ast_g \Bigg] \Bigg\}
	\end{split}
\end{equation}
where $\bm{\ttheta} = \weightv/n$, $\bm{\tk}(\bm{\ttheta}) = \bm{\kp}(\weightv)/n$ and $z^\ast_g$ for $g \in \edges$ are the unique positive solutions of  
\begin{equation}
	z_g \frac{\partial \ln S_i(\bm{z}_i) }{\partial z_g} =  n_0 \ttheta_j   \quad  \forall i \in \{1,\ldots,n_0\}, g \in \edgesci \cap \edgesvj
\end{equation}
We have
\begin{equation}
	\prod\limits_{j=1}^{h_0+n_0}\binom{\liftfac }{ \weightvj_j}^{\dvj-1} \dot= \exp\left\{ n \sum\limits_{j=1}^{h_0+n_0} \frac{\dvj-1}{n_0}H(n_0 \ttheta_j)  \right\}.
\end{equation}
Thus, 
\[
\aiowe_{\alpha n,\beta n} \dot= \sum_{\bm{\ttheta}} \exp(n J(\bm{\ttheta}))
\]
where
\begin{equation}
	\begin{split}
	J(\bm{\ttheta}) = &\frac{1}{n_0}\sum\limits_{i=1}^{n_0}\ln S_i(\bm{z}^\ast_i)\\&-\sum\limits_{j=1}^{h_0
		+n_0}\left[\frac{\dvj-1}{n_0} H(n_0 \ttheta_j) + \ttheta_j \sum\limits_{g \in \edgesvj} \ln z^\ast_g \right].
	\end{split}
\end{equation}
Hence, we have 
$G(\alpha, \beta ) = \max\limits_{\bm{\ttheta}} J(\bm{\ttheta})$
under the constraints 
\begin{align}
	\sum\limits_{j=1}^{h_0} \ttheta_j=&\alpha \\
	\sum\limits_{j=h_0+1}^{h_0+n_0} \ttheta_j=&\beta 
\end{align} 
obtained from \eqref{eq:constrtheta1}, \eqref{eq:constrtheta2}.
Using the Lagrangian multipliers $\mu_1$ and $\mu_2$, the entries of $\bm{\ttheta}^\star= \argmax  J(\bm{\ttheta})$  are the solutions of
	\eqref{eq:solutions_TS_h} and \eqref{eq:solutions_TS_n}, where the values of $\mu_1$ and $\mu_2$ are obtained by enforcing the conditions \eqref{eq:solutions_TS_mu1} and \eqref{eq:solutions_TS_mu2}. \qed

%%%%%%%%%%%%%%%%%%%%%%%%%%%%%%%%%%%%%%%%%%%%%%%%%%%%%%%%%%%%%%%%%%%%%%%%%
%%%%%%%%%%%%%%%%%%%%%%%%%%%%%%%%%%%%%%%%%%%%%%%%%%%%%%%%%%%%%%%%%%%%%%%%%
%%%%%%%%%%%%%%%%%%%%%%%%%%%%%%%%%%%%%%%%%%%%%%%%%%%%%%%%%%%%%%%%%%%%%%%%%


\balance

\end{document}